\documentclass[12pt]{article}
\usepackage{epsfig}

\parskip=\medskipamount 
plus.3em minus.5em \abovedisplayshortskip=.5em plus.2em minus.4em
\renewcommand{\thesection}{\arabic{section}}

\catcode`@=11

\@addtoreset{equation}{section} \@addtoreset{equation}{subsection}
\def\theequation{\ifnum\value{section}=0 \arabic{equation}\ignorespaces
\else \ifnum\value{section}=-1 A.\arabic{equation}\ignorespaces
\else \ifnum\value{subsection}=0
\thesection.\arabic{equation}\ignorespaces \else
\thesection.\arabic{subsection}.\arabic{equation}\ignorespaces
                             \fi
                        \fi
                   \fi}

{\catcode`\'=\active \def'{{}^\bgroup\prim@s}}

\catcode`@=12

\newcommand{\bq}{\begin{equation}}
\newcommand{\be}{\begin{equation}}
\newcommand{\fq}{\end{equation}}
\newcommand{\ee}{\end{equation}}
\newcommand{\bqr}{\begin{eqnarray}}
\newcommand{\beqs}{\begin{eqnarray}}
\newcommand{\fqr}{\end{eqnarray}}
\newcommand{\eeqs}{\end{eqnarray}}

\newcommand{\rf}[1]{(\ref{#1})}

\def\bop#1{\setbox0=\hbox{$#1M$}\mkern1.5mu
    \vbox{\hrule height0pt depth.04\ht0
    \hbox{\vrule width.04\ht0 height.9\ht0 \kern.9\ht0
    \vrule width.04\ht0}\hrule height.04\ht0}\mkern1.5mu}
\def\Box{{\mathpalette\bop{}}}                        

\begin{document}
\thispagestyle{empty}

\vskip .6in
\begin{center}

{\bf Integrability in String Theories}

\vskip .6in

{\bf Gordon Chalmers}
\\[5mm]

{e-mail: gordon$\_$as$\_$number@yahoo.com}

\vskip .5in minus .2in

{\bf Abstract}

\end{center}

The solution term by term to the scattering of all consistent
string theories is given.  The moduli space of M-theory is 
derived and connects the various string theories.  The solutions 
contain both the perturbative and non-perturbative sectors of the 
string.  Modular forms found by differential equations on subspaces 
of the M-theory moduli space and transfinite algebras play an 
essential role in deriving the coefficients.  Various results and 
identities in algebra are found from the explicit solution.  
Archetypes and models are presented in accord with phenomenology 
and cosmology.  

\vfill\break

\noindent{\it I.  Introduction}

String theories in various dimensions have apparently been unified 
with an underlying structure, which arises from M-theory \cite{Witten1}, 
\cite{Witten2}.  The known 
five consistent string theories are Heterotic, IIB, IIA, I, and 
${\rm Spin}(32)$, 
although there is no complete classification until now.  The 
manifestation of the symmetries 
of M-theory, in conjunction with the basic principles of scattering, 
are used in this work to construct the full S-matrix solution to these 
five superstring theories in addition to an (unexpected) five more.  

The full M-theory moduli space has not been presented in the literature.  
Neither has the full solution to the string scattering, including 
the ghost sector.  The full solution means, as a power series expansion  
in $\alpha'$, every prefactor is a non-perturbative function of the 
string model specific couplings.  In this work the moduli space and 
the non-perturbative scatterings are given, and with relatively simple 
differential equations the latter are defined and can be solved.  The 
symmetries and the dualities of the various superstrings  
are both listed and manifested.

\vskip .3in 
\noindent{\it II.  S-Matrices and Effective Actions}
\vskip .2in 

The S-matrix in perturbative string theory is defined formally 
in a couple of ways.  One can use the path integral and its 
world-sheet action, and by gauge fixing all of the extraneous 
degrees of freedom sum over inequivalent random 2-d Riemann 
surfaces.  The amplitudes are defined as the sum of these punctured 
surfaces with wavefunction overlaps provided by the set of vertex 
operators, which are in 1-to-1 correspondence with 
the physical states.  The gauge fixing separates into several well 
known methods such as light-cone, but the integrals are very 
complicated to evaluate by hand.  

In recent years an altered expansion has been developed, which is 
very compatible with the dualities in M-theory.  
The derivative expansion takes the form of the S-matrix and expands 
it as a power series in the number of derivatives and 
operators most of possess 
varying dimension.  This is not a string coupling expansion; the 
field theory form of the 
expansion can be obtained in \cite{Chalmers1}-\cite{Chalmers9}, and 
the string form can be found 
in \cite{Chalmers10}-\cite{Chalmers16}, with some overlap between the 
two.  However, every order in 
$\alpha'$ is explicitly non-perturbative in the string coupling.  As 
$\alpha'$ does not transform under S-duality, the expansion is suitable 
to manifest the duality structures of the string(s).  This expansion 
has been explicitly checked, to find out if there was anything wrong 
with S-duality without a trace.  Supersymmetry and S-duality have been 
presumed independent, which is clearly not true in accord with this work 
and prior work.  The $\Box^3 R^4$ and lower orders have been generally 
investigated together with a few special series, 
and to genus two with the $\Box^2 R^4$ term \cite{Chalmers10},
\cite{Green1}-\cite{Green8},\cite{Berkovits3}.  Subtle issues in the 
genus expansion have been checked off as compatible 
with S-duality \cite{D'Hoker1}-\cite{D'Hoker9}.  These calculations can 
show mechanisms for cancellations in the maximally supersymmetric 
IIB theory and were necessary to show the consistency of S-duality 
with the string and its 
low-energy limit.  In the form of maximally extended supergravity defined 
in field theory there are implications for extended finiteness properties 
\cite{Chalmers10}-\cite{Chalmers11}\cite{Bern1}-\cite{Bern2}.  
  
Consider the graviton scattering in IIB superstring.  Its 
expansion is, in Einstein frame,  

\bqr 
S=\alpha'^{-3} 
 \int \sqrt{g} ~\Bigl[ R+ \alpha'^3 f_{3/2} R^4 + \alpha'^5 f_{5/2} 
 \Box^2 R^4 + \ldots \Bigr] \ ,  
\fqr 
which is power series without end in the number of derivatives.  The 
gravitational portion is shown, but with supersymmetry and the higher 
order terms, including the tensor structure the functional is complete.  

Formally, 

\bqr 
S=\int \sqrt{g} ~ \sum \alpha'^{-3+2j} f_j ~{\cal O}_j \ , 
\label{formalseries}
\fqr 
with ${\cal O}_j$ spanning the basis of operators, which depends 
specifically on the particular superstring theory and its compactification 
including fluxes.  The derivative form is also useful in the 
background field expansion as exemplefied in the holographic 
correspondence.  Determination of the functions $f_j$ then 
'solve' the string theory, apart from resummations which may 
be very relevant in certain energy regimes.  Resummations also 
are expected to show the stringiness of the individual string 
fields such as the graviton $g_{\mu\nu}$, but from the target 
space-time point of view. 

The expansion in \rf{formalseries} is perturbative, but in the 
string scale $\alpha'$.  Variations of the action with respect 
to the fundamental string fields generate the on-shell S-matrix.  
Furthermore, ghost string fields may be included in the operator 
content.  In this work, the coefficients $f_j$ are determined 
based on symmetries of transfiniteness and also the M-theory 
moduli space.  These coefficients, which are functions in the 
couplings, are determined a specified and non-trivial differential 
equation, and the functions $f_j$ are related from one superstring theory 
to another.  

\vskip .3in 
\noindent{\it III.  Transfinite Algebra and Representations} 
\vskip .2in

The affine group transformations appear in the string in a variety 
of ways.  First, they generate the modular subgroup of the Virasoro 
algebra on the world-sheet.  Second, the target space conformal 
representations is expected to relate to the world-sheet Virasoro 
due to the conformal and Virasoro mapping.  Third, there is reason 
to believe that the transfiniteness is related to the integrability 
in non-trivial curved backgrounds such as when black holes are 
present.  For these reasons, transfinite groups and their actions 
are important to use and to clarify their role.  Their use in this 
work involved representation content primarily, without the full 
group and dynamical implications.  There are also powerful connections 
in mathematics, such as in algebra, differential forms and specifically 
equations, and number theory, with the transfinite algebra and its 
interwoven fabric in the full string scattering; this is not explored 
here for brevity.

The affine transfinite algebra $L_{a,b}^\alpha$ is characterized by 
the operations, 

\bqr 
[L_a,L_b]= L_c + \alpha \delta_{a=b} L_a \ ,  
\fqr 
with the Poisson-Lie commutator 

\bqr 
\{L_a,L_b\}=L_a \times L_b = L_c \ ,     
\fqr 
with $c=a\otimes b$.
The affine extension of the Cartan algebra is labeled by the circle 
parameter $\alpha$.  Two representations of $L_{a,b}$ at $\alpha=0$ and 
at once are, 

\bqr 
L_a=e^{i{a \pi \over 2N}} \qquad L_a= {a\over p(a-p)} \ , 
\fqr 
where $p$ is defined as the largest prime factor of $a$; $a\otimes b$ 
are $a+b ~{\rm mod}~ 4N$ and $a\otimes b = p_a+p_2 ~{\rm mod}~ 
C_{p_1,p_2}^{p_1+p_2}$.  A less trivial cyclotomatic field 
representation is generated by the roots to the polynomial 

\bqr 
P(x)=\sum_{\rho=0}^N x^\rho ~{\rm mod}~ p \ , 
\fqr 
with generator and root, 

\bqr 
\alpha_a L_a \qquad \alpha_a = S^{\alpha=1}_{f(N,p),g(N,p)} \ .  
\fqr 
This representation is not used in this work but is important 
to string theory: it species branched covers of Riemann surfaces 
modulo five-form flux charge and conservation or flux conservation 
when a fiber disappears or is damaged.  The cyclotomatic fields 
are generally used in so many areas of string theory that the 
representation is pervasive.     

The commutation relations generate the invariant for the algebra 
$L_{a,b}^{\alpha=1}$, 

\bqr 
[L_a,I]=I  \qquad I=\sum_{-\infty}^\infty \alpha_n L_n 
\fqr  
in which the sum generating the $I$ ranges from minus infinity 
to infinity, which is a bit non-standard.  There is a map that 
may be used to switch between the two labelings.    

In the various string theories the components of the representations  
are the various operators found by taking various products of the fields.  
For example, the operators ${\cal O}_1=R^4$ which is a product of 8 Weyl 
tensors with a specific contraction, or ${\cal O}_2=\psi^4$ the product 
of four fermions 
such as gravitinos.  The appropriate sum of representations span the 
full basis to the analytic part of the effective action after completely 
supersymmetrizing the elements.  The commutator structure is realized as 
a direct product of the two operators, that is, when the operators are 
labeled at the same point.  There are special representations that 
span the basis of the non-analytic operators, such as those required 
by unitarity; string theory constructs can be used to construct these.  

The S-matrix is found by superimposing all of these terms, together with 
the supersymmetric extension, in the form 

\bqr 
S=\int \sqrt{g} \quad \sum \alpha_n L_{a_i,b_i}^{\alpha=1} \ .  
\label{transsmatrix}
\fqr 
The $L_{a_i,b_i}^{\alpha=1}$ is labeled by the set of integers 
from $i=1,\ldots,N_R$, the number of representations.  The 
representations are labeled by the indices $a_i$ and $b_i$.  
The transfinite representation that spans the ghost sector 
allows the action to be taken off-shell, with an unambiguous 
off-shell prescription.

\vskip .3in 
\noindent{\it IV.  Manifestation of Transfinite Symmetries}
\vskip .2in 

The appearance of the transfiniteness of the string scattering 
is not appearant in perturbation theory.  

Consider the following S-matrix as an element in the algebra, 
with generator, 

\bqr 
\prod ~{\rm exp}\bigl( t_j \int {\cal O}_j\bigr) \ .
\fqr 
The generators correspond to conformal maps on the complex plane.  
The S-matrix then is a generator of conformal maps, including curves.

\vskip .3in 
\noindent{\it V.  Moduli Space of the M-theory Web}
\vskip .2in 

The moduli space of the M-theory vacua is a fundamental space on 
and through which dualities act.  These transformations through 
actions on the moduli space transform the various string theories 
into eachother.  For example, in the IIB sector there is S-duality 
which requires the group $SL(2,Z)=SL(2,R)/Z$ acting on a torus.  
This torus is larger in general, and the space follows essentially 
from the quantization of the various string theories.  

Alternatively, determination of the moduli space a priori allows the 
various string 
theories to be interconnected, and solved for, in terms of the transfinite 
representations and modular solutions to a differential equation 
acting on a sector of the moduli space.  The latter two quantities are  
the coefficients $\alpha_n$ and ${\cal Q}_{s}$, which are an integer 
and a modular form of a particular weight.  Both of 
which are found from the moduli space and multiply the operator content 
of the representations involved in the individual string scattering.  Thus, 

\bqr 
S=\int \sqrt{g} ~ \sum \alpha_n {\cal Q}_{s} L_{a_i,b_i}^{\alpha=1} 
\label{scatteringform}
\fqr   
with $n$ and $s$ proportional depending on the theory and funtions of 
the parameter labeling the elements in the representation 
$L_{a,b}^{\alpha=1}$. 

The basic elements of the M-theory moduli space are as follows:   

1) a torus with mapping class group $SL(2,R)/Z$ (or $SL(2,R)/Z_p$)  

2) a circle $S^1$ with group U$(1)$ that models the eleventh dimension 

3) two branched covers of the torus so that a pair of elliptic genus 
three surfaces are constructed with the moduli possibly fixed.

4) a branched cover of a unit $S^3$ with unramified points, i.e. a 
$S^3$ with one point removed or the space $R^3$; this is not part of 
the moduli space but rather used in satisfying a unimodular condition 
with the points.  The points may be resolved or blown up to construct 
the spaces in 1) to 3).    

\hskip .15in The spaces in 1) to 3) make up the the moduli space from which 
the couplings $\tau=\tau_1+i\tau_2$ are found; the gauge coupling is roughly 
the square root of the gravitational coupling required by modular invariance 
or (anti-) holomorphicity.  They transform under the modular group of the 
torus.  In addition they are non-trivial fibers; for example the $S^1$ is a 
non-trivial fiber around the $T^2$ and the two elliptic surfaces are oriented 
in a diameterically opposite fashion and are fibered non-trivially over the 
$S^3$ in a disc like fashion involving only two coordinates.  The disc-like 
fashion follows from one dimension oriented along the $\tau_2$ direction 
and another dimension along a non-intersecting path in $(S^1,\tau_1)$, both 
directions of which have periodicity.  The path in $S^3$ is non-trivial 
because the $S^1$ has a non-trivial fiber over the entire $T^2$.

The construction of the moduli space begins with 12 points on a unit 
n-sphere embedded in 13 dimensional space, with signature specified 
below.  The space has thirteen dimensions as there is split between 11 
dimensions and a torus fiber.  These points are used as unramified 
points placed on the n-sphere after compactifying the plane, so there 
are now 13 unramified points.  The points are chosen in a symmetric 
fashion so that they are equidistant from eachother, and also in a 
manner so that a polynomial of degree 5 specifies the unramified 
surface.  The rotation of the configuration on the n-sphere allows 
one term in polynomial, the fifth, to be dropped through a Poincare 
reduction (fix one point and rotate).  

The degree 4 polynomial 

\bqr 
P(x)=\alpha_1 x^4 + \alpha_2 x^3 + \alpha_3 x^2 + \alpha_4 x + \alpha_5 \ , 
\fqr 
specifies the unramified surface described by an n-sphere with a symmetric 
ordering of 13 points.  This polynomial also describes a torus, but with 
the parameters $\alpha_1=-1/\alpha_5$ satisfying the antipodal constraint.  
This follows from specifying $n=5$ together with the resolution of a degree 
5 quintic in an ambient 13-dimensional space of arbitrary signature with 
the property that the toric is quintic together with its hyperkahler 
structure, which is Pennington's theorem.  Thus the 
polynomial describes a torus after a Poincare resolution. 
One coordinate is free but used to resolve the torus into a five-sphere.  
Due to the Poincare map, there is one ramified point on the surface, with 
all of the previous 13 unramified points taken off the surface into the 
ambient 13 dimensional spacetime.   The torus will be utilized to 
describe the torus of the $SL(2,Z)=SL(2,R)/Z$ and the remaining 13 points 
are used to resolve the moduli space except four which are not necessary.  
These four points are used to close the $R^4$ into a ball in a fiber-wise 
action, which is still a 4-sphere.  

It is necessary that the parameters labeling the 13 points satisfy a 
unimodularity constraint so that the Poincare reduction goes through.  
This is a determinant constraint on 
13 scalar parameters describing the equidistant distance from one point 
to its neighbor using the round metric for example on the unit 5-sphere. 
These 13 parameters specify the configuration of the 13 points.  The 
unimodularity constraint is that  

\bqr 
det(\alpha_i \alpha_j)\neq n/p \ , 
\fqr  
which is an integer mod its largest prime number being an integer also.  
This is automatic if the following theorem due to is 
used:  Consider a round ball with unit radius and determinant unity of 
its collection of n points.  If n mod its reduction to p points is unity 
then the condition of unimodularity holds.  In our case n$=13$ and p$=5$ 
generate unimodularity as they are coprime.  Our condition holds, and the 
Poincare resolution is smooth.  

The torus with one ramified point can accept a point, placing two punctures 
on it.  The remaining 11 points are left off the torus.  One of the 
unramified points on the torus is then bifurcated into a branch cut between 
two bifurcations, such as the location of two spin fields.  Note that 
bifurcating the other point would make the torus a (hyperelliptic) 
genus two surface, although this is not done here; in this interpretation 
the independent $\alpha$ and $\beta$ cycles are correlated with one or 
two supersymmetries in the target space.  The $S^1$ is then 
given a non-trivial fiber over the punctured and branch tori.  This 
is trivial except for a gauge connection which gives monodromy around 
punctures of holomorphy 1, i.e. unramified poles or branch cuts.  The 
fiber is specified by a pair of integers, or a complex number $n_R+i n_R$.  
Alternatively the fiber can be seen as an $S^1$ over the hyperelliptic 
surface with the two independent homology cycles specifying the monopole 
gauge connections.  

Two of the 11 points are resolved into two gauge independent circles 
of radius one.  Radius one is invariant under T-duality, and there is 
no ambiguity.  There is still an unramified point that is hidden because 
the circle closed on itself; imagine the point an ${\cal O}(\epsilon)$ 
distance from the other end of the string and before closing add 
additional points.  Place an additional three unramified points on each 
of the circles and branch them out after rotating one of the points so 
as to complete a torus with an unramified surface and a trivial fiber.
The three unramified branch cuts can be extended to a genus three surface, 
which is elliptic by definition.  

The previous describes the M-theory moduli space configuration, which 
accompanies the ten-dimensional component of spacetime.  In the following 
the compactification of these dimensions is described.    

\begin{figure}
\begin{center}
\epsfxsize=10cm
\epsfysize=8cm
\epsfbox{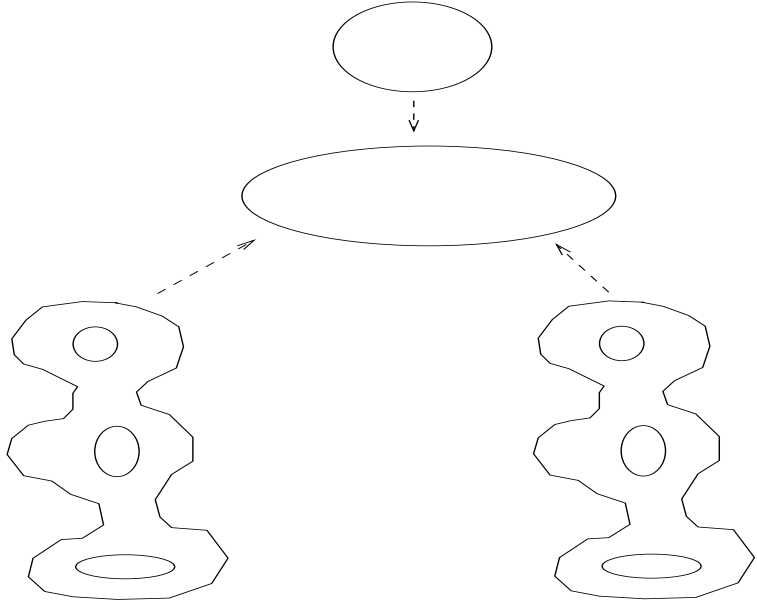}
\end{center}
\caption{Elements of the moduli space.  In the heterotic models an 
additional fiber would be added to represent the gauge degrees of 
freedom.}
\end{figure}

The construction of the moduli space begins with 12 points on a unit 
n-sphere embedded in 13 dimensional space, with signature specified 
below.  The space has thirteen dimensions as there is split between 11 
dimensions and a torus fiber.  These points are used as unramified 
points placed on the n-sphere after compactifying the plane, so there 
are now 13 unramified points.  The points are chosen in a symmetric 
fashion so that they are equidistant from eachother, and also in a 
manner so that a polynomial of degree 5 specifies the unramified 
surface.  The rotation of the configuration on the n-sphere allows 
one term in polynomial, the fifth, to be dropped through a Poincare 
reduction (fix one point and rotate).  

The degree 4 polynomial 

\bqr 
P(x)=\alpha_1 x^4 + \alpha_2 x^3 + \alpha_3 x^2 + \alpha_4 x + \alpha_5 \ , 
\fqr 
specifies the unramified surface described by an n-sphere with a symmetric 
ordering of 13 points.  This polynomial also describes a torus, but with 
the parameters $\alpha_1=-1/\alpha_5$ satisfying the antipodal constraint.  
This follows from specifying $n=5$ together with the resolution of a degree 
5 quintic in an ambient 13-dimensional space of arbitrary signature with 
the property that the toric is quintic together with its hyperkahler 
structure, which is Pennington's theorem.  Thus the 
polynomial describes a torus after a Poincare resolution. 
One coordinate is free but used to resolve the torus into a five-sphere.  
Due to the Poincare map, there is one ramified point on the surface, with 
all of the previous 13 unramified points taken off the surface into the 
ambient 13 dimensional spacetime.   The torus will be utilized to 
describe the torus of the $SL(2,Z)=SL(2,R)/Z$ and the remaining 13 points 
are used to resolve the moduli space except four which are not necessary.  
These four points are used to close the $R^4$ into a ball in a fiber-wise 
action, which is still a 4-sphere.  

It is necessary that the parameters labeling the 13 points satisfy a 
unimodularity constraint so that the Poincare reduction goes through 
smoothly.  This is a determinant constraint on 
13 scalar parameters describing the equidistant distance from one point 
to its neighbor using the round metric for example on the unit 5-sphere. 
These 13 parameters specify the configuration of the 13 points.  The 
unimodularity constraint is that  

\bqr 
det(\alpha_i \alpha_j)\neq n/p \ , 
\fqr  
which is an integer mod its largest prime number being an integer also.  
This is automatic if the following theorem due to is 
used:  Consider a round ball with unit radius and determinant unity of 
its collection of n points.  If n mod its reduction to p points is unity 
then the condition of unimodularity holds.  In our case n$=13$ and p$=5$ 
generate unimodularity as they are coprime.  Our condition holds, and the 
Poincare resolution is smooth.  

The torus with one ramified point can accept a point, placing two punctures 
on it.  The remaining 11 points are left off the torus.  One of the 
unramified points on the torus is then bifurcated into a branch cut between 
two bifurcations, such as the location of two spin fields.  Note that 
bifurcating the other point would make the torus a (hyperelliptic) 
genus two surface, although this is not done here; in this interpretation 
the independent $\alpha$ and $\beta$ cycles are correlated with one or 
two supersymmetries in the target space.  The $S^1$ is then 
given a non-trivial fiber over the punctured and branch tori.  This 
is trivial except for a gauge connection which gives monodromy around 
punctures of holomorphy 1, i.e. unramified poles or branch cuts.  The 
fiber is specified by a pair of integers, or a complex number $n_R+i n_R$.  
Alternatively the fiber can be seen as an $S^1$ over the hyperelliptic 
surface with the two independent homology cycles specifying the monopole 
gauge connections.  

Two of the 11 points are resolved into two gauge independent circles 
of radius one.  Radius one is invariant under T-duality, and there is 
no ambiguity.  There is still an unramified point that is hidden because 
the circle closed on itself; imagine the point an ${\cal O}(\epsilon)$ 
distance from the other end of the string and before closing add 
additional points.  Place an additional three unramified points on each 
of the circles and branch them out after rotating one of the points so 
as to complete a torus with an unramified surface and a trivial fiber.
The three unramified branch cuts can be extended to a genus three surface, 
which is elliptic by definition.  

The previous describes the M-theory moduli space configuration, which 
accompanies the ten-dimensional component of spacetime.  In the following 
the compactification of these dimensions is described.

\begin{figure}
\begin{center}
\epsfxsize=10cm
\epsfysize=8cm
\epsfbox{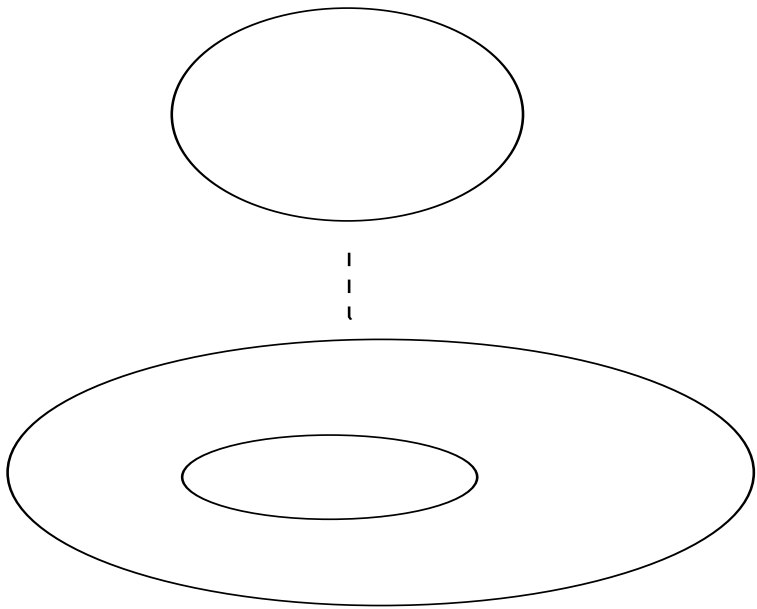}
\end{center}
\caption{The torus of $SL(2,R)/Z$ and the $S^1$ which is fibered 
over it.}
\end{figure}

There are five points remaining, and these are used to define the target 
spacetime theory.  One is used to create an n-sphere by considering 
the n-sphere as an n+1-sphere and setting one coordinate to zero, or 
simply dropping it in a Poincare manner, in a recursive manner, so that 
the m-sphere is obtained.  This is done for an $S^1$, an $S^2$, and an 
$S^3$ using three points.  The fourth space is not as trivial through 
the sphere reduction.  

The four-sphere is obtained by compactifying the four-dimensional flat 
geometry and adding one of the unramified points.  A background background 
$F_5$ flux is added to the sphere to modify it to an anti-de Sitter 
signature.  The point is added to make the metric conformally flat instead 
of spatially flat; the latter is common in the literature.  A point added 
changes the metric to conformally flat instead of spatially flat due to 
a hyperbolic resolvable singularity.  This changes the 
curvature to round instead of flat for a given conformal class, and 
this results in the addition of five flux units thus changing the signature 
to round.  The anti-de Sitter class is then changed to de Sitter, for 
the given amount of five-form flux.  This breaks supersymmetry, but the 
scenario preserves modular invariance as the anomaly in the beta functions 
is still cancelled.

\begin{figure}
\begin{center}
\epsfxsize=10cm
\epsfysize=8cm
\epsfbox{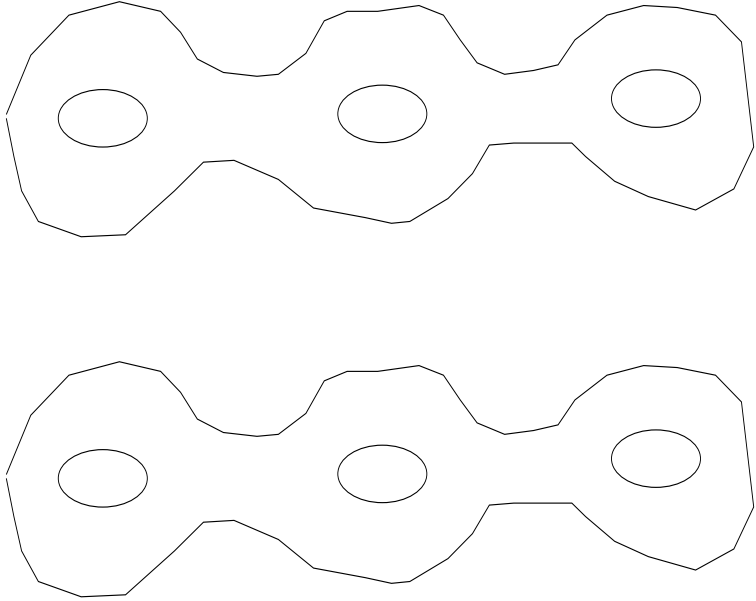}
\end{center}
\caption{The picture shows the unramified extension of the elliptic 
curve into an elliptic Riemann surface of genus three.  The pair are 
diametrically opposite as energy-coherence requires, for they are at 
rest in a stable minimum.}
\end{figure}

\vskip .3in 
\noindent{\it VI.  General String S-matrix and Effective Action}  
\vskip .2in 

The S-matrix of the general string is determined from three 
elements: 

\hskip .15in 1) elements in the transfinite representation 
$L_{a,b}^{\alpha=1}$, with a possible Wick rotation

\hskip .15in 2) coefficients $\alpha_n$ multiplying the transfinite elements 

\hskip .15in 3) modular functions containing the coupling dependence 

\hskip -.15 cm  The coefficients and modular functions in 2) and 
3) multiply the operators that define the S-matrix.  Only the 
analytic in derivative terms are considered, and their form is 

\bqr 
S=\int \sqrt{g}~ \sum \alpha_n F_n(\tau,\bar\tau) {\cal O}_n  \ .  
\label{coeffexp}
\fqr 
The non-analytic terms are also determined through both unitarity 
and special representations of the transfinite algebra.  In fact, 
the computation of the string scattering to all genera and instanton 
number defines these special representations.  

The representations to the string theories are listed in the table, 
\bqr 
 \pmatrix{ 
IIB & & L_{1,1}~L'_{1,2}~L_{1,3} \cr  
IIA & & L_{1,3}~L'_{1,4}~L_{1,2} \cr 
E_8\times E_8 & & L_{1,1}~L'_{1,1}~L_{1,4} \cr 
O(32)/Z_2 & & L_{1,1}~L'_{1,4}~L_{4,1} \cr 
O_1(32)/Z_3 & & L_{3,1}~L_{3,3}~L'_{3,1} \cr 
O_2(32)/Z_4 & & L_{3,3}~L_{3,1}~L'_{3,1} \cr 
O_3(32)/Z_8 & & L_{3,1}~L'_{3,1}~L_{3,3} 
} \ .  
\fqr 
or in a more natural notation, 

\bqr 
 \pmatrix{ 
IIB & & L_{1,2}~L'_{1,3}~L_{1,4} \cr  
IIA & & L_{3,1}~L'_{3,1}~L_{3,1} \cr 
E_8\times E_8 & & L_{1,1}~L'_{5,4}~L_{4,1} \cr 
O(32)/Z_2 & & L_{1,1}~L'_{5,4}~L_{4,1} \cr 
O_1(32)/Z_3 & & L_{3,1}~L_{3,3}~L'_{3,1} \cr 
O_2(32)/Z_4 & & L_{3,3}~L_{3,1}~L'_{3,1} \cr 
O_3(32)/Z_8 & & L_{3,1}~L'_{3,1}~L_{3,3} 
} \ .  
\label{Lfunctionpartial}
\fqr 
The representations and their particle content are further investigated 
in \cite{ChalmersFutPre}.
All of the representations are affine extensions $L^{\alpha=1}_{a,b}$.  
The ghost sector, which contains one tachyon, is marked with a prime 
index.  The left, right representations of $L_{a,b}$ are a gauge 
multipet and a set of supersymmetric partners.  The elements in the 
representations must be supersymmetrized to fill out the entire string 
spectrum; after supersymmetrization the full set of operators is obtained, 
and completely describes the terms in the S-matrix.  The last step 
requires summing over the tower of inequivalent, like Verma modules, 
field contents related to the primary ones listed above; the inequivalent 
ones span the numbers $a$ which differ from those above by multiples 
of $3$.  $b$ is a fixed parameter, which depends on the background 
and its fluctuactions or spatial-temporal dependence.

The matter content of the theories must include the representations, 

\bqr 
E_8\times E_8 & & 3*28*L_{3,1}^{\alpha'=1/2} \cr 
O(32)/Z_2 & & 3*28*L_{3,1}^{\alpha'=1/4} \cr 
O_1(32)/Z_3 & & L_{3,1}^{\alpha'=1/8} \cr 
O_2(32)/Z_4 & & L_{3,1}^{\alpha'=1/16} \cr 
O_3(32)/Z_8 & & L_{3,1}^{\alpha'=1/32} \ ,  
\fqr 
with coefficient $\alpha_n=1$.  The root lattices and the necessary 
fluxes are described in the following sections.  

The corresponding roots associated to the elements in $L_{a,b}^{\alpha=1}$ 
are, 

\bqr 
 \pmatrix{ 
IIB &  1,~\sqrt{2},~\sqrt{2} & & 1,  ~3, ~3  \cr  
IIA &  1,~\sqrt{2},~\sqrt{2} & & 1,  ~2(4),~ 2(1/4) \cr 
E_8\times E_8 & 1,~1,~1 & & 2,~1(1), ~1(5/2)\cr 
O(32)/Z_2     & 1,~1,~1 & & 3,~1(1), ~1(3)\cr  
Type I        & 3,~4,~6 & & 3,~4(2)&      \cr 
O_1(32)/Z_2   & 1,~1,~1 & & 3,~3(4), ~5(1)\cr 
O_2(32)/Z_2   & 1,~1,~1 & & 3,~5(1), ~4(1)\cr 
O_3(32)/Z_2   & 1,~1,~1 & & 
} \ .  
\fqr 
The $\sqrt{2}$ numbers can be changed to $1$ with the following 
modifications: $E_8\times E_8$ has $(1,2,1)$, $O(32)/Z_2$ has 
$(1,2,1)$, ${\rm type} I$ has $(1,2,1)$ and the rest $(1,2,1)$ 
and the set pertains to $\alpha=1/2$ in the affine extension.
These coefficients are stored in the roots $\alpha_n^{a,b}$.  The 
numbers in parenthesis are for spatially varying backgrounds, which 
could be a microscopic effect due to wormholes, or black holes, 
for example.

Apart from these coefficients the functions ${\cal O}_s$ are required.  
These functions are determined in the string setting by an operator 
equation on the appropriate piece of the M-theory moduli space.  In 
the case of the IIB superstring, for example, this operator is 

\bqr 
\Delta_{\tau,\bar\tau}f = \lambda_s f + \prod E_{s_i}(\tau,\bar\tau) \ , 
\fqr 
for an index $s$ partitioned into $s=\sum s_i$, $s_i\geq 3/2$ with 
$s_i$ an integer multiple of one half.  The solution to this equation 
describes the functions ${\cal O}_s$ in \rf{scatteringform}; the 
appropriate coefficient 
$\alpha_n^{a,b}$ multiplies the function, which then multiplies the 
element in the representation $L_{a,b}^{\alpha=1}$.  The different 
string theories have different coefficients, and the latter enter 
into the S-matrix through \rf{scatteringform}.

The modular functions for the various string theories are found 
from the differential equations, or maps, 

\bqr 
IIB \qquad \Delta_{SL(2,R)} f_s=\lambda_s f_s +  
     \sum_{\sigma} \prod f_{s_{\sigma(j)}}  
\label{diffeq1}
\fqr 
\bqr    
IIA \qquad \Delta_{SL(2,R)} f_s=\lambda_s f_s + 
     \sum_{\sigma} \prod f_{s_{\sigma(j)}}  
\label{diffeq2} 
\fqr   
\noindent{$E_8\times E_8$}  
\bqr 
\Delta_{SL(2,R)} f_s +\Delta_{\Gamma(E_8\times E_8)} f_s = 
\Bigl[ s(s-1)+s+{3\over 2}+ \phi\sigma \Bigr] f 
 + \sum \prod E_{s_\sigma(j)} - \delta_{1,s} \ln E_1 
\fqr 
\bqr  
 + \sum \prod E_{w_\sigma(j)}^{8} E_{w_\sigma(j)}^{8}  
 - \delta_{w,1} \ln^2 E_1^{8}  - h(\tau_2)
\label{diffeq3}
\fqr
\noindent{$O(32)/Z_2$} 
\bqr 
\Delta_{SL(2,R)} f_s +\Delta_{\Gamma(O(32))} f_s  
  = \Bigl[ s(s-1)+s+{3\over 2}+ \phi\sigma \Bigr] f_s + 
   \sum_{s_\sigma} \prod E_{s_{\sigma(j)}} 
\fqr 
\bqr 
+ \delta_{0,1} \ln E_{1}(\tau,\bar\tau) 
  + \sum \prod E_{w_\sigma(j)}^{{\rm Spin}(32)} 
 E_{w_\sigma(j)}^{{\rm Spin}(32)}  
 - \delta_{w,1} \ln^2 E_1^{{\rm Spin}(32)}  - h(\tau_2)
\label{diffeq4}
\fqr 
\bqr 
I \qquad  \Delta f_s = \lambda_s^{I} f_s 
  + \sum_{\sigma} \prod E_{s_\sigma(j)} + \sum_{\pm} {\tilde\lambda}_I 
  f_s f_{s\pm 1/2}  \ .  
\label{diffeq5}
\fqr 
The $\Delta$ Laplacian is invariant under $\tau\rightarrow\tau+1$ and 
$\tau\rightarrow -1/\tau$; those $\Delta_{\Gamma(G)}$ are invariant under 
one the lattice groups $E_8\times E_8$ and $O(32)$.  The eigenvalues 
$\lambda_s$ depend on the model.  The derivation of the differential 
equations are deduced in the following sections and require deforming 
the moduli space.

\vskip .3in 
\noindent{\it VII.  IIB S-matrix and Effective Action}
\vskip .2in 

The S-matrix is found by summing all of the 
transfinite representations with the appropriate coefficients, that is 
the modular function and the root number,

\bqr 
 \sum \alpha_n f_n(\tau,\bar\tau) {\cal O}_n  \ . 
\fqr 
${\cal O}$ is an element of a representation; $f_n$ is the modular 
function, and the number $\alpha_n$ is the root label of the transfinite 
representation.  The content is 

\bqr 
L_{1,2}^{\alpha=1},~L_{1,3}^{'\alpha=1},~L_{1,4}^{\alpha=1} \qquad 
  1,~\sqrt{2},~\sqrt{2} \ , 
\fqr  
together with the functions $f_s$ which satisfy the differential 
equation in \rf{diffeq1}.

\vskip .3in 
\noindent{\it VIII.  IIA Superstring Theory}  
\vskip .2in 

The IIA theory is obtained from IIB by T-duality.  Shrink the unit $S^1$ 
to zero radius, causing a conical point.  T-dualize by using an 
S-duality transformation followed by a switch of the following 
transfinite representations in \rf{Lfunctionpartial}, 

\bqr 
 L_{1,2}~L'_{1,3}~L_{1,4} \quad \leftrightarrow \quad 
 L_{3,1}~L'_{3,2}~L_{3,4}  \ , 
\fqr 
in that order; the multiplets are the gravity, gauge, and matter multiplets.  
The point is blown back up to the unit $S^1$.  If the circle is not 
blown down to a point then it is involuted to a circle with radius 
$\alpha'/R$, i.e. the T-dual.

Then the previous modular functions ${\cal E}_s^{IIB}$ 
are replaced to match the appropriate holomorphicity; 
the functions in IIA are the same as in IIB.  The fact that these 
basis functions are identical can be seen from their definition, 
the differential equations \rf{diffeq1}-\rf{diffeq2}.
 
The S-matrix is built from 

\bqr 
\sum \alpha_n f_n {\cal O}_n \ , 
\fqr 
from the representations and coefficients 

\bqr 
L_{1,1}^{\alpha=1},~L_{2,3}^{'\alpha=1},~L_{3,1}^{'\alpha=1} \qquad 
  1,~\sqrt{2},~\sqrt{2} \ , 
\fqr  
and the functions $f_s$ which satisfy the differential 
equation in \rf{diffeq2}. The inequivalent descendants of these 
representations are included in the sum, with their weights.  

\vskip .3in 
\noindent{\it IX.  $E_8\times E_8$ and $SO(32)$ Heterotic Theories} 
\vskip .2in 

The quantum heterotic theory is obtained through the following steps.  
Blow down the $S^1$ and $T^2$, i.e. the $S^3$ Hopf fibration, to a 
double pole.  The fiber structure holds together the configuration 
even when not a double pole from infinite expansion due to the presence 
of matter on the $S^3$.  The origin of the matter is due primarily from 
real states in the ten dimensions made from the fiber bundle of 
n-spheres; the matter has a small but zero probability of quantum 
tunneling onto the $S^3$, and most likely the largest sphere with the 
most matter makes the largest contribution, but this depends on the 
relative position of the 3-sphere with the n-spheres and also with 
the two genus three hypersurfaces.  (General comments about resolving 
singularities in the heterotic theory are found in 
\cite{Cheeger,Silverstein:1995re,HarveyMoore,Kachru:1997rs}.)

The holomorphic (left) half of the IIB ten dimensions should be 
replaced with the holomorphic half of the bosonic string, with 
a 16-dimensional Narain lattice in the $E_8\times E_8$ configuration.  
The $S^3$ still has a two unramified simple poles.  Open the four 
n-spheres, from $n=1,\ldots,4$, by moving the four unramified points on the 
four n-spheres to the $S^3$, one at a time, using the unramified 
pole located at the origin of the $S^2$.   There are five unramified 
poles on the $S^2$ and one on the fiber $S^1$.  To complete the process, 
use the previous procedure to move the ramified pole at the origin of 
the now punctured $S^4$ to the $S^2$, making a total of seven poles 
in total in the $S^3$ fiber bundle.   

Take the elliptic hypersurfaces, and by pinching the holes, change one 
into an $S^2$ with three unramified points (which were created into 
three branch cuts pinned by two poles of order a half).  Then 
the $S^3$ can be extended with another fiber $S^2\times S^2$ over 
$S^2$.  The bundle is an $S^7$, which will admit an Poincare reduction 
to an $S^3$ by setting the $S^2$ coordinates to zero.  There 
are now $12$ unramified zeros which may be removed if the fiber is 
piecewise connected between one $S^2$ and the first $S^2$; it is, due 
to a homotopy condition of Ricci-flatness which is imposed and simple 
to implement.  Consider traversing a closed path from the first 2-sphere 
to the second 2-sphere; this path is homotopy equivalent to a straight 
line, and as such, the same thing can be done in the inverse.  Twisting 
these has no effect, but lifting them to the rest of the bundle could 
have an effect.  The homotopy of any fiber bundle $S^3$ is zero, so 
that there is no obstruction to the assertion of trivial first homotopy.  

Next, three points are removed and used to make 
an elliptic genus three hypersurface, as done in the moduli space 
construction.  Note that there are $9$ unramified points left in the 
7-sphere bundle and one fixed point.  Alternatively, the  
$12$ poles are kept and to ensure that the $Z_7/7$ survives 
in the circle fiber over the 2-sphere when lifted over the $S^2\otimes 
S^2$ the Donaldson-Yau invariant of the fiber bundle must remain as 
is (e.g. s=400,t=40). 

In sum there are two moduli space configurations.  The first is comprised  
of the $S^3$ and two elliptic genus three hypersurfaces.  The second 
is from the elliptic surface and the $S^7$.  Each elliptic surface is 
described by a curve, 

\bqr 
P(x)=\sum^{i=8} a_i x^i \ .  
\fqr 
An involution to an 8-sphere is manifest; the round one has a $Z_8/8$ 
symmetry which becomes $Z_7/7$ when one of the coordinates is set to 
vanish.  Generally there are only 28 
complex structures on the round 8-sphere \cite{Milnor}, and the previous 
sphere specifies only 5 of them.  Choose one 
that is compatible with duality as follows.  Take the round 2-sphere 
and model it with a NUT and 2 parameters.  This is because the elliptic 
surface really has an unramified point that has not become ramified 
yet; the initial curve is not exactly closed when those unramified 
points where placing onto the Riemann surface.  However, before closing 
the torus, which is accomplished by pushing one of the poles onto the 
end of the string, a single Dehn twist is performed.  This causes the 
blown up sphere, after unpinching the hole of the torus and placing 
the unramified point on the sphere, to have a NUT with charge one.  

Having pinched the torus into the sphere, borrow an unramified 
point and put in the center and blow down to remake a torus; Only 
this torus has a NUT.  Push the NUT into the origin where there are 
two poles on top of each other.  The NUT is resolved into a 2-sphere 
with antipodal points identified, which are pushed away from the 
origin so that the solid angle subtended is less than $4\pi^2$.  The 
fiber product $S^2\otimes S^2$ is used to parameterize the complex 
structures of the round 7-sphere.  There are four parameters and a 
radius in the fiber product.  

The round 7-sphere is used as a guide in constructing the exact 
differential that defines the modular functions used in the S-matrix.  
The complex structure must be specified, and it relates to the $E_8\times 
E_8$ complex moduli that enters into its fiber structure over the 
$S^3$ base.  However, the complex moduli are not required for the 
scattering directly, but are required in order to specify which 
branch and the pole location of the differential operator that is 
used to determine the level $s$ functions, such as the ${\cal E}_s$ 
ones in IIB superstring theory.  Choose a direction in the bundle 
$S^2\otimes S^2$ along with a radius and call it $v$; its norm is 
one due to it being defined on a direct product rather than on a 
bundle.  Choose another direction in the fiber product and denote 
it by $w$; its norm is not one and is seemigly equivalent to $v$ 
despite an additional parameter in $w$, but the latter is projectively 
equivalent.  This is clear from the topological index being the same, 
i.e. by Riemann-Roch evaluated at the origin of the root lattice, the 
complex structure has a singularity everywhere but there due to the 
vanishing of the radius.  The vanishing signals that the complex 
moduli jump, but they dont.  The Riemann-Roch identity states that 
the moduli space will jump one unit if there is a singularity or 
branch cut in the complex moduli plane, but only if it cannot be 
removed to another location \cite{Mumford}.  The space has a branch 
from zero on out to infinity due to monodromy of a half, which is 
apparent in the $S^2\times S^2$ if the unramified point is split and 
mapped to a curve in the complex structure moduli plane.  Because 
the point is complex and also unramified, the loophole in the 
Riemann-Roch identity is guaranteed to work.  The topological index 
is the same as the fixed radius to a unit $S^2\otimes S^2$, and 
the number of free moduli is four, including the fixed radii which 
are not allowed to vary.   

The number of parameters in the differential equation require 
the specification of the $SL(2,Z)$ coupling $\tau$ and $\bar\tau$, 
the holomorphic prepontential of the fiber and requires its complex 
moduli to specify without ambiguity, and the lattice of the heterotic 
model.  Consider the following:  Take a sphere and rotate it by one 
unit.  Mark the point and rotate back; then rotate again by another 
unit.  Rotate back and continue until the set of points fills out 
a lattice of points such as the one which $E_8\times E_8$ is described 
by, but in curvilinear planar coordinates.  Two patches are required 
to describe the set of data; however, only one patch is required to 
describe the set of points as the ramified point at infinity does 
not have any data in its (small) neighborhood.  The patch with the 
lattice data can be mapped onto the 2-sphere fiber in $S^2\times S^2$, 
with the exception of a point that does not possess any data locally 
in its neighborhood.  

The lattice data describes the complex structure with a non-trivial 
mapping of a unit disc onto the 2-sphere.  Delete the origin point 
of the disc and map it onto the unit 2-sphere; the anti-podal map 
defines complex conjugation on a section of a $L^2$ line bundle on 
a twistor fibration.  The twistor fibration is required to specify 
a line bundle with the lattice data that can be used to construct 
a holomorphic field with the complex data that is inserted into 
the differential map, in order to define the modular functions.  
Consider the example of a 2-sphere with a line mapped to a region; 
it maps out a line as it moves around, and hence fills a region.  
However it is spinning and some of the time it is multiply covering 
previously covered areas in its region; this is a problem if the 
area it sweeps has to be calculated.  Consider the case of uniform 
rotation and take a pullback to the disk of its rotation and covered 
data; data is black and its covered region is blue, and the lattice 
points are green with a white background.  Take the covered region 
and map it to a second disk, together with the lattice data; the 
twistor information is contained in the set of lattice points that 
are mapped onto the secondary disc.  The secondary disc with the lattice 
locations are then used in a twistor manner to find the time evolution 
of the spinning/nutating line segment.  

In the heterotic case, the unit disc not only contains the lattice data 
but also potential toric information about the blow-up of a Calabi-Yau 
that the lattice could be a fiber over, in the ten-dimensional 
portion of the base space.  This is useful, as the complex line is now 
a curve of cohomogeniety two in the base, but is a line in the fiber 
of $S^2\times S^2$ using only one dimension of the base 2-sphere.  
The fiber $S^2\times S^2$ over the base then describes the time evolution 
of the base manifold, including its complex structure which is what we 
want to find.  The latter can be found as a projection of this multiply 
wound 2-sphere fiber on the base to the real line via a map called the 
Hurwitz construction, which this is partly based on.  The Hurwitz 
construction is too simple when compared with the above; however, it has 
the same feature in determination of a cohomogenous curve from the fiber 
bundle as the Hurwitz map.  The Hurwitz map is based on a twistor bundle 
and a line, and the one adopted here is based on an $S^3$ fiber, a curve of 
cohomogeneity one with respect to the $M_{\rm base}\times S^3$ multiplying 
the line element and cohomogeniety with respect to the latter sphere cross 
line, together with a 
twistor space describing the time evolution of the cohomegenous one curve.   

The twistor space allows us to create a function, given the appropriate 
twistor and lattice data, that reconstructs the fiber if the base manifold 
and its complex structure is known.  Consider a flat ten-dimensionsal space, 
with trivial complex structure.  The curve is not a three-manifold but 
rather a one-dimensional line element by Hodge duality.  The Hodge property 
follows from extending the $S^1$ in the sphere product, picking up the 
curve in the complex plane where the data on the 2-sphere has been stored, 
Hodge-dualizing to a point and then resolving with the aid of the 
Hurwitz-like moment map; the $S^1$ then shrinks back to its former self.  
Else the $S^2$ can be extended as in the $S^1$ case, and this allows for 
a four-dimensional Hodge action to be performed on the 3-volume.  The 
$S^2$ can now be shrunk to its former self, with one exception; this is 
that its conformal class has changed by one unit requiring another unit 
of five-form flux to be added.  Also, the conformal class might induce 
scattering of five-form and gravity with the $E_8$ vectors in a non-trivial 
fashion because it requires instanton-like modes in the gravity sector.  
The instantons are in the ground ring of the $(0,0)$ model  
and their excitations are exponentially suppressed small $E_8$ instantons 
at strong coupling \cite{Witten3}-\cite{Witten4}.  

The Hurwitz-like map can be solved for using the linearization of 
the metric data of the map from the $S^2$ to the complex line.  This 
map, when applied to a sample propagating string, has only two 
excitations, longitudinal and transverse, and it obeys a field 
Monge-Ampere equation in the manner 

\bqr 
\partial_\tau \partial_{\bar\tau} \sigma+ \phi \sigma+\phi^*\sigma= 0 \ . 
\label{hurwitzlike}
\fqr 

The differential equation that the modular functions obey may be 
found directly by reading it off of the fiber bundle structure.  The 
$S^1$ which is fibered over the $S^2$ is also fibered over each of 
the elliptic genus three hypersurfaces.  

The $S^2$ is also fibered 
over each of the two hypersurfaces, and the fiber does not entangle 
with the $S^1$ fiber.  This requires the following three conditions.  
The Chern class of the $S^2$ fiber is trivial, causing the $SL(2,Z)$ 
invariance to be trivially dependent on the auxiliary Riemann surface.  
The fiber should be called auxiliary as its sole function is to 
branch the $SL(2,Z)$ over the elliptic hypersurfaces; this causes the 
non-trivial fiber of the $E_8$ lattice over the elliptic genus three 
hypersurfaces to have non-trivial monodromy around the $SL(2,Z)$ and 
is not trivial.  Last, the $S^2$ over the elliptic genus three 
hypersurfaces can be shrunk to a line, as the fiber is trivial, and 
also stable because only two unramified points are used in its 
overall construction, with one staying at the base of the fiber, 
and 4 are allowed in case points wander on it.  The latter is important 
in the phenomenology as different scenarios require various placement 
of points and depending on the placements the elements in the fibered 
moduli space might be unstable to decay, with a 'pushing' and 'pulling' 
unramified points or the elements might explode.

Each elliptic hypersurface has an $E_8$ fiber on it.  The same $E_8$ 
is also fibered over both the $S^1$, and for compact support to ensure 
the fibers dont 'break', it is fibered over $S_2$.  The configuration 
is also relevant to mass generation of the fermions and is 
discussed in the section on phenomenology.   The total fiber bundle, 
including the elements $S^1$, $S^2$, $M_{g=3}^{d=2}$, $E_8^{(1)}$, and 
$E_8^{(2)}$ are collected in a network of fiber structures; also, of 
the six in the list, the second through sixth items can be made 
holomorphic or anti-holomorphic with three 3 items in each class.  
The holomorphy is useful in that if the number of handles or genus 
is required to change, then it can be doubled from the lowest amount 
by doubling the curve into holomorphic and anti-holomorphic curves.  
The doubling can be used in mass generation and its use is explained in 
the phenomenology section.
 
The various fibers produce the five contributions to a differential 
equation whose solution represents the perturbative and non-perturbative 
contributions to the heterotic superstring.  The $S^2$ fiber of the 
$SL(2,Z)$ has the contribution, 

\bqr 
\Delta_{SL(2,R)/Z} f_s = \lambda_s f_s \ ,  
\label{hetdiff}
\fqr 
with $\lambda_s=s$ because there is no zero of the fiber; it routes 
also through the genus three hypersurface and back to the $S^2$.  If 
the Chern class of the two fibers are the same then the zeros of the 
fiber can be chosen to coincide 
\cite{Cheeger},\cite{Steenrod},\cite{Eilenberg} (these references 
ared throughout in the following).  The chern class of 
the first one is $4$ and that of the second one is $8$; a unit of 
flux on the first fiber raises it to $5$ and lowers the second one 
to $7$ because the unramified point moved.  Add one more flux unit 
and the pair of fibers have flux $6$ and $6$; the flux arrangement 
is a bit unstable because the points want to move around.  Ramification 
of all the points is not advised because particle dynamics in the 
ten dimensional space require some freedom in their placement and movement.    
The points can be ramified but not generally; in this case nail down 
the two unramified points on the $S^1$ and $S^2$ with the ramification 
process \cite{Cheeger} which is not simple as the $S^1$ fiber is not 
trivial.  The ramification points can be brought into the $S^2$ via 
via a hole that is made when an unramified point is pinched off the 
closed genus 0 surface.  This point has to traverse the $S^1$ fiber 
to be able to ramify the unramified point there.  This is possible 
if the fiber has a certain amount of flux to make it stable to 
ramification; this is possible only if the flux is configured even 
temporarily or in the early universe so that the ramification proceeds 
in a stable manner.  The ramification point is then returned to the 
genus three hypersurface and is increased to genus three.  The other 
two points are ramified and unramified with the hole now closed, e.g. 
the job of the former.\footnote{Recall that these two points tell us if there 
is an anomaly in spacetime via an integral of the stress current and 
supercurrent, and is in section 2.}  The flux configuration of $1$ 
unit and $3$ units on the $S^1$ and the $S^2$ fibers; in the 
holomorphic factorization of the fiber, useful in phenomenology, $1$ 
unit of $2$ units on the $P_3$ and the $P_2$ should be consistent with 
duality if the ramifications are chosen to absorb the additional 
$P$ fiber.  This is possible if the ramifications are chosen to coincide 
with the fibers' poles.  Then the $P_3$ and $P_2$ collapse back to the 
former $S^3$ and $S_2$ with the former being a Hopf fibration on the 
$S^2$.  

Much of the previous analysis is addressing the point of whether the 
eigenvalue is $s$ or $s(s-1/2)$.  The latter is more natural, which 
compares with previous modular forms and in previous studies of duality 
of the Heterotic superstring.  There are two points 
on the $S^2$, one of which unramified and now is ramified due to the 
second point.  The two ramified poles on the fiber, which open 
singularities in the base, coincide with the two ramified points there 
and this is reflected in $\lambda_s$ being a polynomial degree two; 
the singularity is brought to Jordan normal form and resubstituted into 
conjugacy class $3$ when it becomes a normal function.  This causes 
the singularity to develop an ankle, or a branch, which stabilizes it 
further; furthermore the fiber develops an infinite class singularity 
when there is a cusp form \cite{Cheeger}.  The fiber is now related by 
duality, $s$ 
to $1-s$ and vice versa; on the keyhole region of $SL(2,Z)$, the duality 
acts this way, and the eigenvalue $\lambda_s$ is $s(1-s)$ up to a constant 
which can be normalized to one by tree-level perturbation theory. 

The $S^1$ fiber contributes the term to right hand side of \rf{hetdiff}, 
\bqr 
 \sum \prod E_s - \delta_{1,s} \ln E_{1} \ .  
\fqr 
The delta function is to cancel the divergence at $s=1$.  This 
contribution is there to make the singularity at the top of the 
$S^1$ bundle go away, and is caused by an unramified point that 
has become ramified by a little more surgery as before.  Then 
the same function as used in the IIB superstring case, which is 
explained now, contributes; there is a singularity at the ramified 
point at the top of the $S^1$ fiber which is nullified by the 
product of Eisenstein functions which is symmetrized due to the 
permutation of the modular ring involved.  Also the singularity 
at $s=1$ is removed by adding the appropriate cancellation term, 
which is an anomalous function in $SL(2,R)/Z_2$, 

\bqr 
\sum \prod E_{s_\sigma(j)} - \delta_{1,s} \ln E_1 \ .  
\fqr 
There is a product with $s_j\geq 3/2$ because the fiber can be 
piecewise be broken up into its smallest and less smallest units 
labeled by $s=n/2$ of positive sign; the permutation group is applied 
because the various combinations of the pieces is totally symmetric.  

The $S^1$ also a combination due to the chirality of the theory, 
that may be represented by a chiral boson somewhere in the fiber.  
It has one fixed point, fixed under the action of chirality, and 
so may be represented on the section (i.e. the modular function) 
by a function contiguous, homologous to its branched covering, to 
the unramified point at the base of the $S^3$.  The function is taken 
to be 

\bqr 
s+{3\over 2}+ \phi\sigma \ 
\label{s1hetcont} , 
\fqr 
where $s$ represents its branched cover, which is a pole of order 
$3/2$ (this is discussed in the section in phenomenology) and is 
chiral because the branch of order $3/2$ has an orientation.  The 
function $\phi$ is a holomorphic function that contributes to the 
metric function as 

\bqr 
g_{z\bar z}= \partial_z \phi_1 + \partial_{\bar z_2} \phi_2 \, 
\fqr 
and has a holomorphic differential.  Here $\partial_z\phi_1$ is 
$\phi-1$, and its abelian differential of the first kind pulls 
out the left term on the right hand side of the above equation 
except for the singularity at the branch cut of the form 

\bqr 
\oint_{P_1} \lambda_{\alpha_1} = \phi_1+\oint_{P_1} \partial_{\bar z_2} 
 \phi_2 = \phi_1+\oint_{P_1} \partial_{\bar z} \phi_2 
\fqr 
because the contour is anti-clockwise and passes over the singularity 
in $P_1$.  This is a branch around the pole in $\phi_2$, the form of 
which is justified by the presence of a zero in the region bounded 
by the contour $P_1$; this means the contribution of the pole cancels 
because the integrand is now anti-holomorphic and the contributon of 
the branch and the zero integrate to zero mod 1 (the loop is the 
butterfly configuration surrounding the zero once and the branch 
twice to give it monodromy 0 mod 2).  The net contribution is 
that in \rf{s1hetcont}.   The function $\sigma$ is genus three 
function on the genus two surface, the torus of duality, which 
has to be present due to the following reasons: it is holomorphic, 
it has two branches due to a double fiber on the $S^2$, it is 
holomorphic in $s$ because each piece of the dissected fiber 
is holomorphic, it has an anomaly at one point because the chiral 
boson can pick up a phase angle around the $S_1$ (the five form 
flux does not couple and there is a deficit of 
charge due to a fixed point at the top of the $S^1$ fiber), there 
is a holomorphic $2$-form on the $2$ sphere coming from the projection 
of the chiral boson onto the 2-sphere and can be 
used to cancel the anomaly from the ramified point of degree $3/2$ 
with a branch extending to another point on the $S^1$.  This 
occurs because the two unramified (one previously ramified) are in 
their most ramified form according to.  There 
a Calabi-Yau is presented that, in accordance with S and T duality, 
encompasses the $S^1$ and is tangential to the ten-dimensional 
spacetime.  This construction reflects this Calabi-Yau though the 
presence of $3$ not $2$ elliptic genus three hypersurfaces; the branch 
of degree $3$ is resolved with three NUTS because of the order $3=3/2+3/2$ 
into a generic but specified genus three elliptic hypersurface which 
can be put together with the other two into a ramified but with circle 
fiber (the eleventh dimension) Calabi-Yau.  It is stable because it has 
a non-trivial fiber with the $S^2$ and the heterotic gauge space without 
any singularities and is Ricci-flat.  The last reason is that the 
removable singularity on the $S^1$ that is unramified can be brought 
to the surface of the Calabi-Yau and the five-form flux cancels its 
presence.  

This results in a contribution of the form \rf{hetdiff} with the function 
$\sigma$ of the form of a Weierstrauss P-function on a genus three
hypersurface that is not hyperelliptic.  This function  
has the form when evaluated at one, thus changing the gauge fiber 
to a ramified one with a branch at the origin and a pole at zero, 
encompassing the entire Calabi-Yau with a curve of cohomegeniety three,   

\bqr 
\sigma=\sum_{k=0}^\infty {1\over k!} k^n \sigma_k \ , 
\fqr 
and 

\bqr 
\sigma_k=P(z_k) \qquad z_k=\tan\bigl(z-2\pi k-{1\over 2}) \ .  
\fqr 
The argument of the Wieirstrauss function is the evaluation at 
all points of the divisor of the reduced hypersurface to a 
hyperelliptic curve of degree $5$.  The moduli are the coupling 
$\tau$ and its anti-holomorphic counterpart $\bar\tau$ when treated 
as a holomorphic counterpart due to the only moduli present.  The 
Calabi-Yau moduli are fixed by requiring that the gauge bundle is 
even and stable over all points.  A fixed moduli can not be 
used in the Wieirstrauss function as there must be singularities 
present reflecting the presence of the points on the which there 
ia a fiber in the base of the $S^2$; there are two points sitting 
on top of eachother and the moduli chosen are $\tau$ and $\bar\tau$.  

The final contribution arises from the gauge degrees of freedom 
fibered over the two genus three elliptic surfaces, which have merged 
into a non-trivial Calabi-Yau.  There are three branched points 
from the holes of the two surfaces, for a total of six; three of 
which close with the other three with branches of order $1/2$ (degree 2).  
The contribution is of the same form as the $SL(2,Z)$ invariant function 
with the $\tau$ parameter holomorphic and the anti-holomorphic 
$\bar\tau$ treated as a holomorphic extension because both are needed 
to specify the anomaly which was cancelled (e.g. the cancellation and 
contribution of the Eisenstein contributions from the two genus three 
hypersurfaces and the $S^1$).  There are no singularities on the fiber 
but there are three branches when represented on the complex plane.  
This suggests that the function contributing has the form, 

\bqr 
\sum \prod E_{w_\sigma(j)}^{8} E_{w_\sigma(j)}^{8} - 
 \delta_{w,1} \ln^2 E_1^{8}  - h(\tau_2) \ ,    
\fqr   
with 

\bqr 
h=\sum_{i=-N}^N \tau_2^{i/2} \ .  
\fqr 
The $E_8$ functions are defined by the root lattice, and the permutation 
is performed as before the fiber can be holomorphically split 
into sub-units of weight number $(w_j,w_j)=(n/2,n/2)$ which are 
then permuted.  The zero mode, or rather the anomalous contribution 
is then subtracted as before.  

The function $h(\tau_2)$ represents the contribution of a single 
degree of freedom which for sake of clarity we call a ghost mode.  
It exists simply because we require a polynomial to ramify over; 
the ramified point's position and type is generated by the polynomial.  
In this case there are three branch points requiring six degrees of 
freedom per point on the Calabi-Yau for a total of 36.  Two more 
specify the orientation of the three branches, as the third is 
oriented diametrically opposite to one of them so the homology 
makes sense in $Sp(3\times 2,Z)$.  The number $N$ is then chosen 
to be $N=19$, $2N=38$.  The anomaly cancellation of the $E_8\times E_8$ 
fiber requires $15$ units of five-form flux to be placed on the $6$ 
points evenly distributed over the points and branches.  One more 
is place on the Calabi-Yau without structure of the points and branches 
so that the $U(1)$ degree of freedom is cancelled there; this occurs 
when supersymmetry is broken.  Without supersymmetry breaking the 
ghost mode cancels and there is an anomaly, so that a tachyon appears 
and is consistent with the theory.  As a result, another transfinite 
representation is included to cancel the modes' oscillation and the 
contribution from $h$.  The tachyon may or may not exist due to 
incomplete or complete cancellation of the function $h$.  The 
coefficients of the function $h$ are taken to unity by moving the 
points and branches inside the Calabi-Yau before ramification.    
Notice that the contribution of the tachyon exists outside of 
perturbation theory as there are contributions with positive 
powers of $\tau_2$, and it is S-duality invariant.  To complete 
the form of the heterotic gauge bundle, take $w=s$.  

There are several scenarios that change the flux organization; this 
is useful for compactifications.  First, instead of flux put on the 
$S^3$, put it on the $S^2$ and arrange for the $S^1$ to be wrap a 
non-trivial 1-cycle rather be embedded in it through the ramified 
points.  Second, instead of flux put on the $S^1$ fiber put it on 
the Calabi-Yau 3 fold around a non-trivial $\pi_1$ generator which 
could be ramified by one of the three oriented branch points; they 
are oriented because the resolution of a triple singularity requires 
that they have oriented directions.  Third, the de Rham complex 
could be used to make the $p$-forms accept the flux; this is 
straightforward as the Calabi-Yau is made up of three fixed moduli 
Riemann surfaces with two non-trivial fibers.  Four fibers are redundant 
except for a branch cut due to having two fibers on the same surface 
which is resolved by adding more flux; more flux might destabilize 
the resolved the singular $R^4$ and make it unstable for cosmology 
which is the case and is discussed indirectly in the later section 
on phenomenology.  The amount of flux is two units and only $4$ 
more can be placed on the singular $R^4$; this would cause the 
singularity to collapse to a point and the big bang would occur 
far sooner than what we expect(ed).  Some flux can be placed on 
the $E_8\times E_8$ $16$-dimensional holomorphic root space, but 
the total allowed strength can not exceed the first Chern class 
which means four units per $E_8$ branch.  $8$ units would compensate 
for the problem of the big bang occuring when it might occur(ed), 
but the extra degrees of freedom from $1$ to $8$ points might 
make gauge phenomenology difficult.  Each flux unit produces a 
naive factor of a $U(1)$ which has to be broken by further 
modification of the compactification and the arrangement of flux, 
and generically $4$ flux units on each branch are allowed.  This 
scenario is not preferred from a gauge anomaly point of view, and 
also from a stability point of view as the closer to unstable the 
configuration is the more modification to a branched but non-singular 
$R^4$ becomes, and also the more the vacuum would produce particles 
such as gauge modes, ghosts, and/or tachyons.  
 
The final issue is the modification of the index on the modular 
functions $E_w^{E_8\times E_8}$.  As observed in 
there is an obstruction to using $w=n/2$ as an index and the one 
$w=n/4$ or $w=n/5$ seems to be liked.  There is an issue with 
embedding the Calabi-Yau in the manner done in  
which was not branched well.  The paper has several branches 
emanating from the Calabi-Yau and based on five branched and 
unramified points at the top of the fiber.  If they are ramified 
then the fiber would be too singular to accept compact support, 
or have functions exist as a result; only $2$ or $3$ ramifications are 
allowed as the Chern class is $7$ for this fiber with an $S^2$ 
emanating from each branch or sub-fiber.  A branched fiber 
requires three units, one each for the point, the branch (or a 
half) and the base point, and as a result only two or three 
of these branches can be held down by the ramifications.  If 
they are not ramified, the fiber will presumably be unstable 
with the branches reverting back to their unbranched form.  The 
Chern class contributes to the energy, and the minimal energy 
configuration is the previous.  

There is a way to avoid the singularity.  Branch the five points 
on a cylinder with two fiber points deleted or a cone with one 
fiber point deleted.  The five points have base the cone and extend 
to the set of branch points on the Calabi-Yau.   Two or one of the 
points have been resolved into spheres with the use of the fiber points, 
and they are ramified.  Two more points are found diameterically 
opposite to the ramified points as the branch points have been allowed 
to pass through the holes and ramified opposite to the holes; they're 
not ramified until a base point together with its branch accepts one 
unit of flux, further flux of one unit is added each to the points opposite 
to the holes.  The holes can now accept two more points, which make 
them immediately ramified as they are holes and no flux is necessary 
to make them stick.  One more ramified point requires $3/2$ or $2$ units 
of flux to make it ramified anywhere on the cylinder or cone; the 
ramification of a point on a cylinder or a cone requires $1$ less 
than than usual due to a gauge field present, which is turned on 
by the five-form wrapped around a non-trivial homology generator; 
the flux necessary is $3/2$ for the ramified point in the Calabi-Yau 
and $1/2$ unit for the base point at the top of the fiber, which 
is total not $5/2$ or $3$. The total flux is then $6$ or $7$, depending 
on whether a branched point requires $1/2$ or $1$ unit of flux to 
ramify\footnote{There is a discussion of this in,
\cite{Cheeger}.  The preference for $1$ unit of flux is used in 
this work.}  Seven units of flux are not allowed on the fiber 
due to the Chern class of $3$, which allows $2n+1$ for stability.   
Then this configuration is not allowed; however, if the other point 
of view of $1/2$ per branch point is used, then the configuration 
is $6$ and hence is meta-stable.  The fiber is so close to unstable 
that it will probably cause particles to be emitted from the vacuum, 
such as gluons, $E_8$ fermions, and maybe tachyons.  This would cause 
further damage to the fiber and it might decay.  

The differential equation generating the modular functions is, 
\bqr 
\Delta_{SL(2,R)} f_s +\Delta_{\Gamma(E_8\times E_8)} f_s = 
\Bigl[ s(s-1)+s+{3\over 2}+ \phi\sigma \Bigr] f 
 + \sum \prod E_{s_\sigma(j)} - \delta_{1,s} \ln E_1 
\fqr 
\bqr  
 + \sum \prod E_{w_\sigma(j)}^{8} E_{w_\sigma(j)}^{8}  
 - \delta_{w,1} \ln^2 E_1^{8}  - h(\tau_2)
\label{typeHetdiff}
\fqr 
with $\sigma$,$\phi$, and $h$ defined earlier.  $E^{8}_w$ is an 
analog to the Eisenstein function on $SL(2,R)/Z$.  The 
contributions in \rf{typeHetdiff} are from five (sub-)fibers.

The S-matrix is then generated by the sum, 

\bqr 
\sum \alpha_n f_n {\cal O}_n \ .  
\fqr 
The representations and coefficients are  

\bqr 
L_{1,4}^{\alpha=1},~L_{1,1}^{'\alpha=1},~L_{1,4}^{\alpha=1},~ \qquad 
  1,~1,~1, \ , 
\fqr  
and the functions $f_s$ which satisfy the differential 
equation in \rf{typeHetdiff}.

\vskip .3in 
\noindent{\it X.  Type I and Type I' Theories}   
\vskip .2in 

In the type I theory the two unramified points on the $S^3$ are collected 
and placed at the origin of the $S^2$, thus producing a singularity of 
order two.  The point at the origin of the $S^4$ in the compactification 
is pulled off and placed on the torus; the presence of the pole of 
order $(1,1)$ signals a bolt $S^2/Z_2$ at the origin of $S^2$.  Blowng up 
the singularity generates an additional $S^2$.    

The $(1,1)$ pole can be split into a pair of branch cuts with endpoints 
of the type, in pairs, $({1\over 2},{1\over 2})$.  There are two 
orientations, $(1,0)$ and $(0,1)$, which require labeling the branch 
cuts with an orientation.  The unramified double pole can be resolved 
into these branch cuts if the orientation is preserved.  The branch cuts 
generate the double cover of the 2-sphere.   

The fiber of the $S^1$ is specified by its behavior in the proximity 
of the two branch cuts on $S^2$, or rather, on $CP^1$.  There are four 
contours generated the homology, of which three are independent.  Call 
the line integrals of a 1-form around one of the branch cuts on the 
two sheets $\alpha_\pm$; the two sheets are labeled by $+$ and $-$.   
The line integral between two points of order $(0,1/2)$ and $(0,1/2)$ 
on different branch cuts is called $\beta_{\pm}$.  Only three of the 
contour integrals are independent and $\beta_-$ is eliminated.  

The integrals 
\bqr 
{1\over 2}(\alpha_+ + \alpha_-) (\alpha_+,\alpha_-) \sim 
 f_s (f_{s+1/2},f_{s-{1/2}})
\fqr 
are in correspondence with the two functions $f_s f_{s\pm {1/2}}$ with 
$s$ an integer multiple of a half.  The gauge bundle is specified by 
integrals of the connection around the contours, and result in two 
integers $m,n$ given the $\alpha_+$ and $\beta_+$ contours.  As long 
as these integers are non-vanishing the branch cuts can be moved around 
on the 2-sphere.   

The differential operator specifying the modular functions is almost 
the same as in the IIB or IIA theory, but due to blowing up the bolt 
it is modified to, 
 
\bqr 
\Delta_{SL(2,R)} f_s = \lambda_s f_s + 
   \sum_{\sigma(j)} \prod E_{s_{\sigma(j)}} 
   + \sum_{\pm} \tilde\lambda_s f_s f_{s\pm {1/2}} \ .  
\label{type1diff}
\fqr 
The sum $\pm$ represents the sum over the two independent contours 
used to label the fiber, the latter of which requires two half-integral 
numbers.  The eigenvalues are $\tilde \lambda_s={1/3}$ and 
$\lambda_s=s(s-1)$.  Cusp forms with a residue at the tip of the  
keyhole region, at $\tau={1/2}+i{\sqrt{3}/2}$, satisfy the differential 
equation $\Delta f_{\rm cusp}=\lambda' f_{\rm cusp}$ (e.g. $s=0$), and 
this normalizes the coefficient $\tilde\lambda$ to $1/3$.  The 
eigenvalue $1/3$ is really a normalization.  For really large values 
of $s$ the first two terms appear to dominate due to the number of 
partitions of $s$ into a set of smaller numbers $s_i$; the fiber term 
$\sum f_s f_{s\pm {1/2}}$ can be neglected; the $SL(2,R)/Z$ moduli space 
requires the eigenvalue $\lambda_s=s(s-1)$.        

The S-matrix is built from 

\bqr 
\sum \alpha_n f_n {\cal O}_n \ , 
\fqr 
from the representations and coefficients 

\bqr 
L_{}^{\alpha=1},~L_{}^{\alpha=1},~L_{}^{\alpha=1} \qquad 
 3,~4,~6 \ , 
\fqr  
and the functions $f_s$ which satisfy the differential 
equation in \rf{type1diff}.

\vskip .3in 
\noindent{\it XI.  SO(32)${}^{(n)}$ Heterotic Theories}
\vskip .2in 

The $SO(32)$ superstring theory follows closely with one 
exception, the gauge bundle $E_8\times E_8$ can be replaced 
with one of the four gauge groups: ${\rm Spin}(32)$, ${\rm Spin}(32)/Z_4$, 
${\rm Spin}(32)/Z_8$ or ${\rm Spin}(32)/Z_8$.  The latter four can be 
exchanged for a lattice model once the points are identified 
on the fiber with a branch cut singularity extending from one 
point to another.  The ${\rm Spin}(32)/Z_8$ has two forms as 
the Cartan sub-algebra can be modded in several ways and two 
of them are inequivalent.  They have rank $2$, $4$, $8$, $16$, 
and $32$.  

The inequivalent representation of the weight lattice is arranged 
to have the diametrically opposite Cartain generators in the 
${\rm Spin}(32)/Z_8$ algebra to be on a line or at an angle 
of $60$ or $600$ degrees; the latter is on the $6$-fold copy 
of the realization of the weight lattice on the complex plane.  
The twisting of the gauge degrees of freedom require a specification 
of the five-form flux which is consistent with the remaining 
flux configuration and also modular invariance together with S- 
and T-duality.  This is accomplished by turning on the five-form 
flux such that $2$ units are applied to each resolved point 
on the gauge fiber and $2$ units spread over the entire fiber.  
$2$ units are applied to the resolved singularities so that 
the NUT that they contain can be converted into a $S^2$ sphere 
with a punctured hole; there is a branch cut emanating from the 
ramified point diameterically opposite representing the Cartan 
generator which passes through the hole to its modded partner 
Cartan generator.  An additional $2$ units of flux is all over 
the fiber so that it wont be unstable, as by Cheeger's theorem 
an equal amount of flux on ramified points together with some 
flux on the entire fiber can be stable only if the difference, 
i.e. the Cheeger quantity, is $6$ or $12$ on a branched 
two-dimensional surface; in this case there are $8$ Cheeger 
numbers and $6$ or $12$ with a plus or minus sign are required 
for $16$ dimensions.  As there are $16$ Cartan generators in 
the sub-algebra, the Cheeger number must be $6$ or $12$ mod 
$n$, with $n$ the number of ramified points.  There are $32$ 
ramified points and all are branched; thus a flux of $2$ mod 
$32$ is required.  The first Chern class prefers $2$ in order 
to be stable.  

The embedding of the $SO(32)$ algebra into ${\rm Spin}(32)$ 
allows for the possibility of a heterotic duality between 
the points and the algebra.  There are sixteen resolved NUTs 
and one resolved bolt that each contain a $S^2$ of complex 
structure.  One of which is prefered, which is the bolt.  
Take the bolt and cover it with a $S^2$ so that the hole 
is patched; this is accomplished with one unramified point 
used in the construction of the bolt resolution aftwerwards 
the bolt becomes ramified.  The bolt can be rotated and its 
covered hole oriented to any other point in the fiber, using 
the Fubini-Study metric on the fiber base extended to act on 
the entire gauge bundle.  Then there is an $SO(30)$ action 
on the points due to a sequence of orderings together with a 
shift of the complex phase as directed for by the bolt; it is 
non-abelian due to the fact that the sphere is curved and the 
orderings of directings do not commute.  This group fills out 
an $E_8\times E_8$ model as the phase angles must fill out a 
representation of a maximally extended $SU(2)$ which is either 
${\rm Spin}(32)$ or $E_8\times E_8$.  The $SU(2)$ is a 3-form 
which projects as a two-form on the $S^2$ but with the remaining 
component specifying one of the angles in the $S^2$; their 
is no redundancy as the radius has not been set to one.  The 
actions of the rotations must match with the Cartan sub-algebra, 
which is non-trivial due to commutation relations; however, there 
is a subtlety in the construction as not all points exist on 
side of the bolt which could be anywhere.  Fix it to be at the 
end of the gauge fiber.  The spinning bolt can label the Cartan 
generators with a number and the remaining generators are 
resolved singularities with ramified points, which are all numbered.  

The structure constants are defined to reflect the inherit curvature 
of the $S^2$ and the location of the generators, one of which is 
$H_1$ the prefered Cartan generator used to mod the ${\rm Spin}$ 
algebra by $Z_p$.  The $Z_p$ modding reflects the generators by 
adding or subtracting $2\pi/p*N$ to the rotation curve between 
one point and the other; $N=2$ for $8$ generators, $N=4$ for 16 
generators, $N=1$ for $4$ generators.  Each curve sweeps out a 
trajectory in the $S^2$ with arclength equal to the phase angle 
with the modding $\rho\rightarrow \rho\pm 2\pi/p*N$, and $N=1$ 
is special because it has no elliptic cover, i.e. a branch of 
the $S^2$ is not required to uniquely specify the path in case 
the curve runs back on itself.  The prefered generator $H_1$ is 
required not to do this.  The $H_1$ requires a commutator 
with itself, and the line element could go around the sphere 
and come back to itself making a mod $p$ cover of the trivial 
traverse to a point and then traverse back.  This construction 
is iterable, and the line elements form a group once the structure 
constants are defined.  The line elements close an infinitesimal 
distance apart, proportional to the curvature of the 2-sphere.  

The group is either $SO(32)/Z_p$ or $E_8\times E_8$ due to the 
following; the generators on the fiber are rank $16$, and those 
of the line elements in the $S^2$ are $16$ with the possibility 
that some duplicate or copy other elements' actions.  The group 
generators might roll in back of another group generator if 
the fiber is unstable, which appears not to be the case because 
the first Chern class is $0$ mod $N$ on a manifold of dimension 
$16$; this amount of charge is not normal and the bundle could 
unwrap \cite{Cheeger}.  The bundle may not only unwrap but rather 
it might dissipate if further electric charge enters, which means 
that it is not wanting of further charged particles like the 
electron, real or virtual.  As the bundle is relatively stable,  
the gauge group is constructible and makes sense.  

The gauge group must be $E_8\times E_8$ and this is the ${\rm Spin}/Z_p$ 
theory.  It is rank $16$, is maximally extended from $SU(2)$, and 
it has a bilinear form of rank 2, inherited from the $3$ form which 
can integrate on the $S^2$.  The latter is degenerate over the patched 
hole on the $S^2$ as it doesnt accept any more charge; it should be 
blown up into a surface that the $2$-form could integrate over.  The 
bundle is then an $S^4$ with two two-dimensional surfaces forming the 
direct product space without fiber; as such the product of the two-form 
can be used as an integration element and this cannot be a form as the 
infinitesimal differentials each appear twice but in an organized fashion.  
Each $SO(32)$ generator has a direction, if they dont roll.  First, if 
they roll, then the Cartan sub-algebra must be relabled.   Second, the set 
of 2-spheres might roll into an $E_8\times E_8$ gauge group from the 
point of the bolt $S^4$, which still has a $S^2$ fiber; this is the 
basis for the duality.  One can form an $SL(2,Z)$ invariant fiber 
shockwave that causes all the balls to move in particular directions 
to capture the $SO(32)/Z_p$.  

All of the balls are now well-ordered, but one of them, the Cartan 
generator has rolled to a new position behind another ball.  This 
process occurs until all the balls are either behind or in front 
of their individual sets.  The group depends on the configuration  
of the sets.  In order for the gauge group to be maximally extended 
the balls must line up in an $E_8\times E_8$ setting, which takes 
time without the specific shockwave applied at the base of the $S^2$ 
bolt; there is a temporary fiber from the $S^2$ of the $SL(2,Z)$ to 
the bolt and requires one unit of flux to pin down the ramified point 
at the top of the fiber.  There is only one unit because the $S^2$ 
at the bottom of the fiber requires only one unit and the gauge 
bundle is sufficiently rigid that no charge is necessary; here rigid 
means that it is close to unstable and naturally unramified points 
are broken and branched without need to ramify.  
The shockwave is instantaneous.  

The configuration rolls into an $E_8\times E_8$ configuration and T-duality 
allows one to unroll the balls for a moment to analyze their configuration.  
T-duality takes the complex structure of the bolt and exchanges it with 
the K\"ahler structure of the bolt, which is close to zero, and might even 
vanish.  To keep it from vanishing a Taub-NUT singularity is added that 
makes the curvature far from zero; of course the singularity has to be 
added to cancel but for the moment it is present.  To keep the singularity 
from almost vanishing, the bolt requires an almost complex extension, such 
as a Newijnhuis tensor without compact support.  This makes the balls 
rotate backwards as time has no direction without support.  If they rotate 
backwards then they will eventually roll back into their present position, 
and this can be facilitated by another quantum well such as that induced 
by a small and better focused shockwave.  Then they are stuck in the heterotic 
position by applying five-form flux that rigidly holds the configuration.  
Thus T-duality has been allowed to map $SO(32)/Z_p$ to $E_8\times E_8$ 
by the flow of the balls from the exceptional heterotic to the orthogonal 
heterotic and then back by T-dualizing, and waiting for the balls to 
roll, to the $E_8\times E_8$ theory.  T-duality will work if the fact 
that the K\"ahler modulus and the complex structure modulus are interchanged 
is recognized; in this case, there are four allowed configurations: ${\rm 
Spin}(32)/Z_p$ with $Z_p=1,2,4,8$.  Because there is one hole and one 
singularity on the ball now, the ball sees two objects simultaneously 
one behind the other and is required to do so.  That limits the group 
cases to the ones mentioned with the balls in each row related to each 
other by the quotient.

Now that the group action is identified, the balls must rotate 
into their respective position to fill out the vector or adjoint 
representation.  In $E_8$ there is no adjoint but the vector takes 
its role, so that the adjoint may not be so trivial.  The representation 
follows from symmetrizing the balls in their current position in each 
of the sets and calling the first ball their respective Cartan generator; 
the remaining balls in the set are descendents.    

The fiber structure is almost the same between the heterotic orthogonal 
case and the exceptional heterotic theory, with one exception.  Because 
there are now five theories, the lead candidate must be chosen to 
supervise the remaining candidate groups.  This means that total supervision 
of the tower of heterotic orthogonal groups is under the control of the 
heterotic orthogonal $Z_2$ group, and that once in this phase a 
simple transformation can alter the theory to any of its descendents.

The flux configuration is nearly identical in all of these theories with 
one exception.  Instead of $2$ fluxes on the gauge bundle there are 
$2$ mod $N$ with $N$ being related to the Cheeger sum.  That is, the 
sum of all the flux units on the gauge bundle must be equal to $2$ 
modulus the sum of all the flux units on the unramified points to make 
them stick so that their combination mod $N$, the number of flux units 
on the points, is equal to zero.  See the section above for further 
details.  This is the primary distinction in the model(s) from the 
exceptional heterotic case.   Another distinction is the amount of 
flux on the fibered Calabi-Yau; in this case, as there is fewer flux 
on the gauge bundle the flux can be increased in two ways on the $3$-fold.  
One attach it to points on the manifold with branched covers, which have 
been glued and resolved.  Second, attach it to the overall manifold without 
increasing the sum beyond the Chern class.  Further alternatives are not 
mentioned.  Third, a single point can still be added to the $S^2$ in 
the $S^3$ fiber bundle; this breaks S-duality as the quotient is now 
$SL(2,R)/Z_2^2$ which is not a group but rather a point-like object.  
The singularity can now accept four to eight more units of five-form 
flux without destroying the black hole singularity at the origin of the 
$R^4$ space.  This could be useful for model building.  

\vskip .3in 
\noindent{\it XII.  Compactification and Supersymmetry Breaking}
\vskip .2in 

The compactifications considered here go against lore in one 
aspect; primarily anti-de Sitter spacetimes are considered, 
but locally de Sitter ones are presented that presumably model 
our neighborhood and local galaxies.  Cosmological data cannot 
distinguish a redshift with positive or negative cosmological 
constant if the blueshift is treated as the microwave background 
anisotropy.  This scenario fits with rings of compression separated 
by hundreds of millions of lightyears and as a shockwave of sorts 
in the early universe, near the time of big bang nucleosythesis.  
The latter assertion would explain why there is little signature 
of the blueshifted matter and would be found primarily in the CMB 
dust.  Also, our galactic halo could be considered red-shifted 
if the compression wave entered our galaxy sooner than expected, 
for example in a few hundred million years with an additional 
compression wave explaining some of the blue-shift that data 
has observed in the halo; alternatively a compression wave 
may have entered our galaxy later such as in the billions of 
light-years.  It is difficult to use current data to indicate 
either scenario, the given above or the one with a matter dominated 
universe with uniform expansion rate and constant cosmological.   
The expansion rate is consistent with the Hubble's constant if 
the compression waves are placed apart and have the right movement 
outward from our point of view, and if Moore's law is obeyed.  

\vskip .3in 
\noindent{\it XIII.  IIB Phenomenology}  
\vskip .2in 

Remove the point from the $S^2$, after moving to the $S^1$.  The 
point then becomes ramified at the origin of the $S^1$ due to the 
non-trivial fiber.  This point can be unramified by taking in a 
point at infinity and then sending it back.  The unramified point 
on the $S^1$ is used to open the $S^1$, then moved to the a 
four-dimensonal subspace of the ten coordinates.  Alternatively, 
a point on one of the elliptic surfaces can be unramified by 
pinching it off of one of the closed holes and used to unramify 
the one on the circle fiber.  The fiber trivializes as the 
genus drops from genus $3$ to genus $2$; if not trivial then 
the 3-form can be used to make the fiber almost trivial, and giving 
it a vacuum expectation value.  The vacuum expectation is going 
to modified anyway and it is assumed to be not the right value 
for phenomenology.  

The point is taken off the end of the line segment and placed on 
a four-dimensional subspace in the ten dimensions.  Because it 
placed by hand it is ramified, and can be blown up as a NUT of 
degree 1 into an anti-de Sitter space.  

The $3$-form is taken off the $S^1$ fiber where it was used to 
trivialize the fiber and moved onto the ten-dimensional space.  
The resolution of the NUT singularity requires five units of 
flux, which gives the appropriate number to cancel the target 
space gravitational beta function and restore modular invariance.  

There are various scenarios in the following.  Close the remaining 
five coordinates so that a $S^3$, $S^2$, and $S^1$ are made each 
with a point removed.  First construct an $S^3$ fiber over the 
$S^2$, with the two singularities moved on top of eachother; $\pi_3(S^2)$ 
is the set of integers and conformal class of one is required for 
modular invariance due to the absence of non-integer spins.  One point is 
ramified due to the non-trivial fiber 
and one is unramified, both of which are located at the base of $S^2$.  
Then treat the two $S^1$ circles equally which means fiber both of 
them over the $S^3\times S^2$ bundle; $\pi_1(S^3\times S^2)$ is 
the set of integers by Kunneths formula, including the resolution 
of the double pole at the origin.  The unramified hole is on top 
of the ramified hole and both are resolved into a crosscap with a 
double charge located at its antipodal points.  $\pi_2$ of the 
crosscap is the set of even integers and contributes not to the 
Kunneth formula except for a double charge divided by two.   The 
crosscap is then a part of the $S^3$ fiber and can be closed by 
the addition of another two blown up points.  

There are two $S_1$ fibers over the $S_3\times S^2$ bundle.  The 
fibers are each specified by an integer as the $S^3$ fiber is 
specified by a complex number, one for the Chern class and one 
for the conformal class evaluated on the tangent bundle for the 
latter.  The 
flux is chosen to cancel that on the anti-de Sitter space.  Two 
five forms can be taken together as $F_5\wedge F_5$ and given 
the vacuum expectation values.  Indeed, if a triple product is 
taken, and the singularity on one of the genus 3 elliptic hypersurfaces 
is blown up appropriately, then the third form can be used to compensate 
for the flux in all directions so that they add to zero; this occurs 
if these two hypersurfaces, the $S^1$ fiber in the eleventh direction, 
and the first ten dimensions are taken as a product, this gives 
a manifold interpretation to the presence of the flux and its 15-form.    
The fluxes are chosen to agree with phenomenology.  

In a different 
scenario, the amount of flux used to alter the bolt at the origin 
of $R^4$ can be altered by adding and subtracting no more than 
two units as the bolt can withstand only these values before 
requiring the additional resolution of a new singularity; the 
time scale of the instability is presumed shorter than the age 
of the universe.  The remaining flux is spread among the $6$ 
compactified dimensions with flux: one unit on the circle, two 
on the genus three elliptic hypersurface, and three on the $S^3\times 
S^2$.  This requires four units on the bolt located at the origin 
of the $R^4$ which is now resolved in a de Sitter space.  

Another scenario is $4$ on the elliptic hypersurface, $2$ on the circle, 
and $4$ on the bolt.  Another is $4$ on the elliptic hypersurface, 
$2$ on the circle, and $4$ on the bolt except one unit is stolen 
from the $S^1$ fiber with a puncture so as to make a total of $5$ 
on the bolt; this is anti-de Sitter but in order to flip the sign 
of the $5$ ten units are given to the $S^1$ branch singularity which 
it now becomes due to the large number of flux units and its 
instability (its instability is localized to a branch singularity 
which is the most general form of a point-like singularity of degree 
one).  

(One more noteworthy solution is the ramification of the 
singularity caused by the ten units of flux and its placement on the 
$2$-sphere which fills its hole, else it can be used to partially 
compensate for the pinched hole on the elliptic hypersurfaces.  The 
presence of the point on the hypersurface might cause the entire 
Riemann surface to become unstable after blowing up the pinch to 
genus three.  If it becomes unstable then it might pinch the 
$S^2$ fiber above the ten dimensional space to a point (which is 
zero coupling) and then also collapse the remaining genus $3$ elliptic 
hypersurface.  This continues until only the four-dimensional 
space survives, which will provide the solution is anti-de Sitter, 
and causes it become a de Sitter space with $4$ or $5$ units of flux.  
This scenario has a large order pole at the origin of spacetime.)   

The presence of holes in the $S^1$ and and a required bolt filling-in 
in the $S^3\times S^2$ fiber together with the application of the 
bolt resolution requires three points to be brought in from infinity 
and one returned.  Depending on the accuracy of the phenomenology 
this scenario could be physical.  Two points could also be effectively 
removed from the genus three elliptic hypersurfaces.  The stability 
of the configuration might be drastically altered, and is not prefered, 
as discussed in the section addressing this matter and its signature 
in cosmological data.  

\vskip .3in 
\noindent{\it XIV.  Heterotic Phenomenology} 
\vskip .2in 

The configuration of the IIB superstring allowed for two supersymmetry 
breakings, either to $(0,0)$ or $(0,1)$.  The heterotic breaking is 
identical to the IIB superstring, with the exception that two points 
(one ramified and one unramified) are required for a pole of order 
$6$ and only one ramified point is required for a pole of order $8$ 
\cite{Cheeger}; 
these are located at the origin of the $R^4$.  A higher degree 
singularity with unramified points are less singular than expected 
because the resolution is facilitated by their movability.  The 
number of allowed flux units varies, from $1$ to $4$ in the former 
and $1$ to $9$ in the latter, leading to a wider 
class of supersymmetry breakings.   M-theory suggests that the 
flux configuration would remain close to the same in the Heterotic 
phase as in the IIB superstring phase.  A computation of the groundstate 
energy of the vacuum is required to determine the rate and possibility 
of a different string phase is one corner of the universe.  

The mass and the couplings can be found by analyzing the singularity 
at the origin of the $R^4$.  Examine the resolution of the heterotic 
pole at the origin.  A NUT of type 1 is blown up into a $S^2$ with 
two points removed, suggesting a cylinder between them, i.e. $S^1\times S^1$.  
The full blow up requires three points to be consistent with duality in a 
holomorphic setting; this results in $S^1$ times $S^1\times S^1$ with a 
complex coordinate spanning the latter, a fiber of Chern class $3$ as 
there are two unresolved blow up points and a ramified point at the 
origin of the former $S_1$.  The ramification may not be removed as it as 
located at the origin and the fiber is non-trivial; more points could 
be brought in from infinity but this is not availaible as in the 
superstring case only one point from infinity was required and for 
consistency this condition is unaltered.  

In \cite{Chalmers17} the masses of the known particles were found to 
high accuracy, satisfying the mass formula \cite{Chalmers18}, 

\bqr 
\Lambda \bigl({\Lambda\over m_{\rm pl}}\bigr)^{n/16}\Bigl[1 + 2^i 5^j + 
  ({\Lambda\over m_{\rm pl}})^{-3} + \ldots  \Bigr] \ , 
\label{massformula}
\fqr 
which is accurate to 1
fits the pattern with another suppression of $(\Lambda/m_{\rm pl})$.  
For some reason there are $27$ combinations of numbers $2^i 5^j$ between 
$1$ and $1000$ with the scales $1,10,100,1000$.   

The masses can be obtained in accord with the mass formula in 
\rf{massformula}.  Take a bilinear fermion pair and move it onto 
a holomorphic $S^1$ fiber after passing the initial $S^1$; it completes 
a loop via the anti-holomorphic $S^1$ fiber.  A phase could be added 
to the fermion bilinear, as described in \cite{Chalmers1}, that acts as  
$\bar\psi\psi\rightarrow e^{-n/16} \bar\psi\psi$; the phase could also 
be interpretated as that of the dihedral group in $SO(4,2)$ which suggests 
that another four points are added to the $R^4$ and resolved.  This 
action was identified in \cite{Chalmers1} as an orbifold of a $S^4$ by 
$\Gamma_{4,2}$, but an orbifold of an $S^5$ with group action from 
$\Gamma_{4,3}$; $\Gamma_{4,2}$ is $D_4$ and $\Gamma_{4,3}$ is $D_{11}\wedge 
D(\Gamma_{5,1}$ and they have dimension $8$ and $19$ \cite{Cheeger}.  
Three of the latter are not close to the origin and are not used; 
the branch point which is unstable due to nine points on top of eachother 
sits in between the origins of the two group actions.  

The $D_{11}$ group is good for conformal points when all the points sit 
on top of eachother, but resolved.  The remainder acts when the points 
are separated and also resolved.  Each action leaves the bilinear 
unchanged except one if the centralizer $H_1$ is chosen to be in the 
corresponding Lie algebra.  The phase angle is $n/16$.  There is no 
ambiguity as the axis, the Cartan generator $H_1$, is prefered.  However, 
the other group actions are complicated but the leave the phase unchanged.  
 
Next, the $\Gamma(5,1)$ is treated as a conformal extension of the $D_{11}$ 
known as Kummer's algebra  
and permutes the six dihedral vertices amongst eachother.  This action 
changes the phase on each vertex as the wavefunction passes due to the 
action of the Cartan generator $H_1$ which permutes the phase angles 
at the vertices.  The actions are chosen in an oriented pathwise fashion 
from the following steps: 1) choose which vertex the wavefunction is 
entering, 2) permute its phase angles by $H_1$, 3) chose another vertex 
to hop to, 4) permute its phase angle, 5) repeat until the wavefunction 
exits though the point it entered after hopping each vertex-vertex only 
once, which is unique.  The phases allowed from hopping are: 
$e^{n/5}$, $e^{n/6}$, $e^{n/7}$, $e^{n/8}$, $e^{n/9}$, $e^{n/10}$ except 
for one which is resubstituted with $m\rightarrow m+1$ due to angle 
deficit.  The remaining four generators possess angle deficits from 
$30$ degrees to $90$ degrees in increments of $10$ degrees, with a 
$\pm 10$.  

Due to the uniqueness of the path through every vertex, the phase angles 
must add to $p/16$ with $p=1,2,3,4,5,6$ \cite{Chalmers4}.  The quarks 
and leptons fall into this category.  When a neutrino enters the chamber 
point split the first point in the finite lattice to include a branch and 
give it a phase angle of $n/4$ with the other vertices unchanged.  
The sum of these terms is $n/16-5$; the factor of $5$ occurs due to the 
Cartan generator $H_1$ acting on too many vertices at the same time.  In 
this case $H_1$ has an action on the entire lattice each time a node is 
reached.

With the phase angle attached to the fermion bilinear, the pair 
exits the region by traversing outward from the resolution of the points.  
Imagine a black hole, in the early universe, and the fermion pair  
traverses through the horizon from the inside outward (or because there 
is no casually disconnected region because the fermion could exit through 
a path in the higher dimensions).  It acquires a mass proportional to the 
phase angle, with a scale set by $\Lambda=1$ TeV and $m_{\rm pl}$ so that 
the first term in \rf{massformula} is found.  This is alluded to partly 
in \cite{Chalmers17}.      

\vskip .3in 
\noindent{\it XV.  M-theory in higher dimensions}  
\vskip .2in 

The natural number of dimensions for M-theory is thirteen; the moduli 
space presented contain a fibered set of three 
dimensions with a ten dimensional spacetime.  
There are ten in the usual superstring description, and three more 
in the moduli space; there is a circle fiber over a 2-sphere.  The 
strong coupling limit of IIA is obtained by an S-duality transformation 
followed by the limit $\tau_2\rightarrow\infty$.  The 2-sphere in the 
IIA configuration is a 2-torus with complex structure $\tau$, and its 
volume goes to infinity.  The two torus is the same as a sphere with 
a puncture at one point and an unramified point at the antipodal point; 
this is due to pinching off and unramifying the point in the middle 
of the torus as the modulus hits the cusp point at $(1,\sqrt{3}/2)$ 
thus generating a 2-sphere with a hole at one end and an unramified 
point at the other.     

One limit is as follows.  Recall that the volume $V_S$ of the circle 
satisfies $V_S=\sqrt{V_{T^2}}$, the volume of the torus.    
The strong-coupling limit is via $g_s\rightarrow \infty$ which shrinks 
the torus to a point on which the circle is fibered over.  The unramified 
Rpoint is moved onto the circle and splits into a line segment.  The line 
segment can be extended to fill the entire real line, and together 
with the ten dimensions, is eleven dimensional spacetime. 

A second limit is as follows.  The unramified point exists at a location 
away from the fiber on the circle, which is then equivalent to a line 
segment after it splits open using its ramification the circle into 
two pieces as it has to be branched.  The singular points are absorbed 
by the branches and the two pieces are glued on top of each other, which 
removes the branches.  The volume to infinity of the punctured 2-sphere 
is the complex plane and the fiber to the circle becomes trivial in this 
limit.  This results in a thirteen dimensional spacetime.  

A conformal field description can be given to the eleventh dimension 
which is compatible with eleven dimensional supergravity.  Put a spin-1, 
spin-2, spin-3/2, and spin-1/2 field with $2^{d/2}$ components on the 
circle which also contains the singularity of order two.  The degenerate 
limit completes the eleven dimensional supergravity.  A five form is 
added with one component fibered over the eleventh branch cut, due 
to disingularizing the two unramified points into an eleventh 
dimensional branch with a spin field at $x_{11}=\pm 1$ and another 
one in the tenth coordinate at $x_{11}=\pm 1$ which is oriented in 
the opposite direction.  Desingularizing the non-trivial fiber of the 
five-form on both branch cuts requires fixing the fourth and fifth 
components on top of the branch, losing its field content.  The three 
form is then used to fill out the topological degrees of freedom in 
the supergravity lagrangian.  The five-form is relevant to carry over 
its and other degrees of freedom via supersymmetry to the IIB theory. 

The two theories in $11$ and $13$ dimensions can be given consistent 
conformal field descriptions.  In the first case there are ten flat 
spacetime dimensions and a compact circle of infinite radius fibered 
over the eleventh 
coordinate, with an unramified point added to the circle.  The point 
at infinity is unramified and stolen; together with the unramified 
point at the origin, a singularity of order two is sitting at the origin 
of the eleventh dimension.  

Rather than this description, the point is ramified to the origin of the 
eleventh dimension as the circle fiber is non-trivial.  As the circle fiber 
is taken to have infinite radius, the point at infinity is stolen, unramified, 
and moved on top of the ramified point at the origin of the tenth dimension.  
Together the two points resolve a Taub-NUT class A singularity with a bolt 
and a nut, thus unramifying the formerly ramified point.  The two unramified 
points are moved back onto the circle and placed at the origin.  As the 
fiber becomes non-trivial the coordinate completes the flat space eleventh 
dimensional spacetime.  Both this limit and the one in the previous 
paragraph lead to the same result.

In the thirteenth dimensional example a conformal field model can be 
presented based on the black hole work \cite{Chalmers18} together with 
that in \cite{Witten1}.  Consider a black hole wound around the $S^1$ 
with a $U(1)$ (charge) fiber extending outward, in the other non-compact 
direction.  T-duality changes these the $U(1)$ fiber with the other 
non-compact direction, which can be thought of as a wormhole due to a 
singularity at the origin of the $S^1$; this is due to the non-trivial 
$\pi_1$ and the winding of the soliton solution and its fiber.  This 
blackhole and its wormhole relative have a description in terms of 
a holographic massive topological conformal field theory.  Consider 
a WZW model at level $1$ and its central extension; the former is 
described 

\bqr 
S=\int d^3x ~\varepsilon^{\mu\nu\rho} W_\mu W_\nu W_\rho + 
 \int d^2x\sqrt{g}~ g_{\mu\nu} W^\mu \partial^2 W^\nu \ , 
\fqr 
with $W_\mu$ a spin field transforming in the $1/2$ representation.  
The central extension is given by the usual mass term in three 
dimensions, 

\bqr 
S=\int d^3x ~\varepsilon^{\mu\nu\rho} W_\mu W_\nu W_\rho 
 + \int d^2x\sqrt{g}~ g_{\mu\nu} W^{\mu} m^2 W^\nu \ .  
\fqr 
Both terms are included to preserve a transcendental symmetry so 
that upon $m^2=-k^2$ the action of the combined terms vanishes.  
This configuration describes a self-dual wormhole that vanishes 
at the radius by momentum conservation.  

For consistency, the compatibiity with eleven dimensional supergravity 
is explained.  Throw away the topological term; then dimensionally 
reduce the spin field $W_\mu$ (recall that in 2-d, spin $1/2$ is 
equivalent to spin $0$).  The mass term has $m^2=2\pi n$.  Take $n=1$, 
which labels this sector as the $d=11$ supergravity limit one.  Absorb 
the $2\pi$ by a field redefinition; this is possible in another sector, 
but not all sectors simultaneously.  The spin reduction generates the 
field content $(2,3/2,1)$ with the counting $(1,2,1)$ of the degrees 
of freedom.   Take the reduced spin field and branch it over the 
two Riemann surfaces with the spin $\pm 1$ over the individual components. 
As the two surfaces are equal except for being diameterically opposite, 
project one to the other and mutiply the components by a factor of two. 
There are eight homology generators on the elliptic genus three 
hypersurface.  This means that in light-cone gauge the gravitational 
multiplet in $d=10$ should extend to $d=11$, and by supersymmetry the 
Noether current is subtended by a factor of three for dimensions, 
which become flat at large radius except for small corrections.  This 
explains the origin of the field theory limit in from the $d=13$ corner.  
The three-form arises by multiplying the individual components to make 
a three-form or by Hodge duality an 8-form by wedging all the components 
together; this shows that the topological term is not fundamental but 
rather a consequence of duality involving the three-form on the $S^3$, 
which is homotopically equivalent to a three-sphere.  The duality 
extends holomorphically to the product of the three-sphere together with 
the $d=2$ Riemann surfaces, which becomes a 7-sphere.  The eight components 
of the spin field span the individual components subject to a constraint 
that light-cone tells us is spin independence, i.e. two independent spin 
structures generate the same spin content.  This condition is reflected 
in the topological aspect of the three-form, as the former does not require 
a metric to distinguish the spin dynamics on the $S^7$ in a strong sense.  

The fibered $S^3$ has a maximum invariance without the topological 
sector of $E_{4}$.  The topological sector includes two copies of 
the graded Lie algebra $E_1$, which possesses a transfinite representation; 
the $Z_2$ breaks this down to $E_1$, which is interesting as the $E_4$ 
has two Cartan generators, enough to specify the indices $a$ and $b$ 
in the representation content $L_{a,b}$.  Upon extending the $S^3$ 
fiber bundle to encompass one non-compact dimension such as the tenth, 
the $E_{4}$ becomes an $E_6$ broken down to $E_5$ because there is 
no point at infinity.  The $E_5$ when taken as a semi-direct product 
with the single exceptional Lie algebra $E_1$, can be maximally 
extended into the gauge symmetry algebra of $E_{11,11}$ and is 
non-compact and isomorphic to the 3-ball.  This is the purported 
maximum symmetry of M-theory, however, it is theoretically possible 
to extend further by including the remaining $E_1$ which was projected 
out by a discrete homomorphism.  In doing so, the symmetry algebra 
becomes $E_{13,13}$, which has not been constructed yet, if possible; 
the $Z_2$ becomes one of the generators in the extension and in the 
surface parameterization, thus restricting its action to a hypersurface.  
This should be the maximum symmetry, as pertinent to the work here.  

Reshuffling the indices on the representation labels, the eigenvalues 
of the two Casimirs of $E_4$, can lead to a group theoretic understanding 
of the representation numbers of $L_{a,b}$.  However, once this is done 
there is a complication in the transfinite math that leads to the 
labeling in the first place analogous to reshuffling the spin 
representations of the Lorentz group which is simpler when the Poincare 
group is used in conjunction with the Pauli-Lubanski operator.  Here 
$(a,b)$ are two Casimir eigenvalues of the $E_4$ and take on values 
which are integer multiples of a number in the string theory application.  
The numbers are ordered into $a>b$ by a similarity transformation which 
has action on the pairwise path integral and coherence path integral, 
together with the infinite tower of representations following the triples 
described earlier.  This matrix is the Fubini-Study matrix of the 
transfinite extension of the $S^2$ without altering its complex structure; 
transfinite here means that several points were added to the $S^2$, without 
altering its metric or complex form (after deleting a point) 
\cite{Hammermesh1}.  This means that the matrix can be represented on 
the plane without boundary as a complex number; this also means that there 
is a non-linear field redefinition of string theory so that the 
representations enter Jordan normal form, or that the matrix is diagonal.  
The Jordan normal form can be thought of as a reduction of the classification 
of the representations $L_{a,b}$ if one thinks of the transfinite 
operations as conformal transformations in the complex plane, as 
discussed in the first section.  Having only diagonal elements 
indicates that there is no central charge and thus no Weyl anomaly in 
the critical superstring, but rather from the target space-time 
point of view; this is also in the presence of three extra dimensions, 
which would naively signal a $c=3$ anomaly except in the $E_8\times E_8$ 
and $O^{(n)}(32)$ models, which would have an $c=6$ anomaly.  
The Jordan normal form could also simplify the dynamics even in 
non-trivial backgrounds. 

The $S^1$ fiber must contain a chiral boson \cite{Siegel1}  
in order for the transfinite 
symmetry to have a central extension apparently.  This chiral boson 
transforms like a scalar under the action of Poincare translations 
and as a Lorentz multiplet of spin $(1/2,0)$ under Lorentz transformations 
in $d=11$.  The scalar seems to be required as the $d=2$ sphere is 
holomorphically equivalent to the plane with the point removed and 
placed on the 1-sphere as an unramified point; the topology suggests 
that the radius is the affine parameter $\alpha$ and the chiral bosons 
partition function when interpreted as a conformal field theory enhances 
to an $SU(2)$ thus generating a conformal block with three primary 
towers.  These towers generate the indices of the $L_{a,b}$ in an 
$SL(2,Z)$ invariant manner without any obstruction due to the chiral 
nature of the partition function; there is no anomaly if the point 
is placed at the origin, and the partition function is a scalar with 
a factor of $1/2$ multiplying the modes.  This is T-dual to the 
coefficient of a $4$ in the mode expansion.  \footnote{In type IIB a similar 
partition appears with a factor of $2$ and generates the full 
contribution to the four-point function, with similar 
partition functions describing higher-point amplitude contributions.  
However, a result due to 
Hardy-Littlewood states that these partition functions are 
equal if the mode oscillators are normalized with that of the partition 
function of the one with a factor of $2$ or $1$; this appears 
trivial.  

Apparently, if the 
tangent space of the $S^1$ plus a point has its tangent space replaced 
with the tanget space of an $S^1$ so there is no holomorphy then the 
IIB partition function in \cite{Chalmers13} is obtained, and this 
signals the transfiniteness of the IIB superstring without the circle 
fiber, which has been reduced to a point so that string modes cannot 
propagate on it.}  The factor of $4$ is useful in transfinite 
number theory as it is often associated with the last non-transcendentally 
solvable polynomial system, i.e. a quartic.  It is conjectured that 
the expansion of the partition function into a series, as in 
\cite{Chalmers13}, will generate all string related transfinite 
representations.  The factor of $4$ signifies that there are 
three contributions, the gauge and gravity, the ghost, and the 
matter components; also the Hardy-Littlewood result stating that 
one can replace $4$ with $2$ prevents the numbers in the series 
expansion to grow without bound and has practical application 
in the computations (even replacing $4$ with $1$ appears reasonable 
if the computation is such that the terms in the series contributing 
to a specific $x^N$ are collected appropriately). 

\vskip .3in 
\noindent{\it XVI.  Summary and Outlook} 
\vskip .2in 

The superstring and its moduli space have been described with the 
goal of elucidating its M-theory relation as well as its exact solvability.  
In all cases, the exact S-matrix is constructed in terms of transfinite 
representations and certain moduli forms that satisfy a differential 
equation on a corner of moduli space.  This work has simplified the 
approach to scattering so much that now it is possible to simply write 
down the S-matrix; this is advantageous for practical computing.  

Phenomenology of the superstring, primarily in the IIB and exceptional 
heterotic, is described.  The required five-form flux has to be distributed 
in the compactified spacetime and moduli space; this places constraints 
on the scenarios, some of which are described.  The higher dimenional 
spacetime of $S^1\times S^2\times S^3$ appears to be one of the simplest 
realistic configurations, including possible quantization of the radii 
due to the quantized but ordered amount of flux on the circles.  This 
scenario is improved by the resolution of a 
singularity at the origin of the $R^4$, and this has implicatons for 
cosmology.  In this stable compactification the realistic mass value 
generation of the fundamental fields such as fermions are given.    

It is possible to interpret the moduli space as a higher dimensional 
inclusion to the ten dimensional spacetime.  Thirteen dimensions are 
required as the moduli space is three beyond the critical superstring 
dimension.  There are a few bubbles of empty space that exist in this 
three-dimensional world, and they are partly populated with strings 
that quantum tunnel back and forth between the ten-dimensional and the 
three-dimensional space.  The stability of this configuration, which 
requires the quantum coherence effect, is 
excellent and can be used as a further guide in finding the exact 
orientation of the three-dimensional configuration in various phases 
of string theory; vacuum selection is eliminated by including the 
coherence part of the energy.  WMAP data appears to indicate that 
the above picture is correct \cite{ChalmersToAppear}.  

The exact form of the scattering in any non-trivival theory usually 
has impact in other fields, including mathematics.  The exact solution 
of all of the superstring theories, in uncompactified spacetime and 
where its clear, in compactifations such as tori or Calabi-Yau manifolds 
with toric moduli spaces, should have similar impact.  Indeed, 
there are dozens of results following from solving the superstring theories.  

\vfill\break

\end{document}